%% file: sample-authordraft.tex
\documentclass[sigconf]{acmart}
\AtBeginDocument{%
  \providecommand\BibTeX{{%
    \normalfont B\kern-0.5em{\scshape i\kern-0.25em b}\kern-0.8em\TeX}}}

\copyrightyear{2023}
\acmYear{2023}
\setcopyright{acmlicensed}\acmConference[CHI '23]{Proceedings of the 2023 CHI Conference on Human Factors in Computing Systems}{April 23--28, 2023}{Hamburg, Germany}
\acmBooktitle{Proceedings of the 2023 CHI Conference on Human Factors in Computing Systems (CHI '23), April 23--28, 2023, Hamburg, Germany}
\acmPrice{15.00}
\acmDOI{10.1145/3544548.3580953}
\acmISBN{978-1-4503-9421-5/23/04}

\usepackage{subcaption}
\usepackage{csquotes}
\usepackage{multirow}
\usepackage{enumitem}

\newcommand{\h}[1]{\textbf{#1}}

\newcommand{\ftest}[3]{$F$(#1,#2) = #3}
\newcommand{\pvalue}[1]{$p $ \textless #1}
\newcommand{\pvalueequal}[1]{$p$ = #1}
\newcommand{\pvaluegreater}[1]{$p$ \textgreater #1}
\newcommand{\etasq}[1]{$\eta_{p}^{2}$ = #1}
\newcommand{\msd}[2]{$M$ = #1, $SD$ = #2}
\newcommand{\corr}[1]{$r$ = #1}

\begin{document}

\title[What Shapes People's Reactions to Algorithmic Harm]{Blaming Humans and Machines: What Shapes People's Reactions to Algorithmic Harm}

\author{Gabriel Lima}
\email{gabriel.lima@kaist.ac.kr}
\orcid{0000-0002-2361-350X}
\affiliation{%
  \institution{School of Computing, KAIST \& Data Science Group, IBS}
  \country{Republic of Korea}
}

\author{Nina Grgi\'{c}-Hla\v{c}a}
\email{nghlaca@mpi-sws.org}
\orcid{0000-0003-3397-2984}
\affiliation{%
  \institution{Max Planck Institute for Software Systems \& Max Planck Institute for Research on Collective Goods}
  \country{Germany}
}

\author{Meeyoung Cha}
\email{mcha@ibs.re.kr}
\orcid{0000-0003-4085-9648}
\affiliation{%
  \institution{Data Science Group, IBS \& School of Computing, KAIST}
  \country{Republic of Korea}
}
\renewcommand{\shortauthors}{Lima \emph{et al.}}

\begin{abstract}
    \input{content/0abstract}
\end{abstract}



\maketitle

\section{Introduction}
\input{content/1intro}

\section{Background}
\input{content/2background}

\section{Blame Judgments of AI Systems, Developers, and Users}
\input{content/3studyIntro}

\subsection{Study 1}
\input{content/4study1}

\subsection{Study 2}
\input{content/5study2}

\subsection{Study 3}
\input{content/6study3}

\section{Meta-Analysis}
\input{content/7meta}

\section{General Discussion and Implications}
\input{content/8implications}

\section{Concluding Remarks}
\input{content/9conclusion}

\begin{acks}
This work was supported by the Institute for Basic Science (IBS-R029-C2) and the National Research Foundation of Korea (RS-2022-00165347). 
\end{acks}

\bibliographystyle{ACM-Reference-Format}
\bibliography{sample-authordraft}

\input{content/10appendix}

\end{document}

%% file: content/0abstract.tex
Artificial intelligence (AI) systems can cause harm to people. This research examines how individuals react to such harm through the lens of blame. Building upon research suggesting that people blame AI systems, we investigated how several factors influence people's reactive attitudes towards machines, designers, and users. The results of three studies ($N$ = 1,153) indicate differences in how blame is attributed to these actors. Whether AI systems were explainable did not impact blame directed at them, their developers, and their users. Considerations about fairness and harmfulness increased blame towards designers and users but had little to no effect on judgments of AI systems. Instead, what determined people's reactive attitudes towards machines was whether people thought blaming them would be a suitable response to algorithmic harm. We discuss implications, such as how future decisions about including AI systems in the social and moral spheres will shape laypeople's reactions to AI-caused harm.

%% file: content/1intro.tex
Artificial intelligence (AI) systems assist and make decisions in many high-risk scenarios (e.g., medical diagnostics~\cite{esteva2017dermatologist}, bail decision-making~\cite{propublicastory}, hiring~\cite{nytresume}), and their increasing use has raised new ethical concerns. Machines have been shown to discriminate against racial and gender minorities~\cite{obermeyer2019dissecting,reutersamazon} and even caused the deaths of their users~\cite{nytteslapassengers}. Understanding how humans react to the harms caused by AI systems is critical for determining how these systems should be governed, developed, and adopted~\cite{bonnefon2020moral,cave2018portrayals}. Past research has shown that people underreact to algorithmic discrimination~\cite{bigman2019holding}. However, the opposite has also been reported, as evidenced by citizens attacking autonomous vehicles following the death of a pedestrian by a self-driving car~\cite{nytknives}. These contradictory stories highlight the difficulty in comprehending people's collective reactions to algorithmic harm.

The current research is motivated by this challenge. We examine people's reactive attitudes\footnote{We use the term ``reactive attitudes'' to refer to people's judgments of blame.} towards algorithmic harm through the lens of blame. There is ample empirical evidence suggesting that people blame AI systems and robots when they cause harm~\cite{lima2021punish,malle2014theory,furlough2021attributing,kim2006should,kahn2012people}, and numerous philosophical discussions have centered on whether machines are appropriate subjects of blame~\cite{tigard2021artificial,champagne2021mandatory}.

We present three studies ($N$ = 1,153 in total) that investigate how various factors influence laypeople's reactions to algorithmic harm. We first explore whether AI systems providing simple explanations about their decisions affects how much people blame them, their designers, and their users. Scholars have argued for explainable AI systems as a way to hold humans accountable when machines cause harm~\cite{robbins2019misdirected,jobin2019global}.
In Study 1, we empirically test this proposal in the context of algorithmic decision-making and find that providing simple explanations has no effect on blame judgments. Nonetheless, blame directed at developers and users increase when the AI system presents a clearly discriminatory explanation for its decision (e.g., based on gender or race). The same effect does not occur in the case of machines, implying that people may underreact to algorithmic discrimination when it comes to assigning blame. 

An exploratory analysis of Study 1 identified two factors that correlate with how people attribute blame to AI systems and other actors: 1) the perceived harmfulness of the decision and 2) the participants' attitudes towards AI systems. In Studies 2 and 3, we manipulate these factors and find that blame judgments of developers are largely shaped by the perceived fairness of the algorithmic decision, whereas blame directed at users is influenced more by its perceived harmfulness. A meta-analysis of the three studies indicate that fairness and harmfulness have little to no effect on blame towards machines, which is instead strongly associated with people's perceptions of AI systems as blameworthy agents.

Both developers and users are blamed highly for algorithmic harm, implying a joint responsibility framework in which people direct their reactive attitudes towards multiple actors~\cite{hanson2009beyond}. We discuss how laypeople's expectations may influence future regulatory decisions, which should consider the possibility of those in power exploiting explanations to shift perceived responsibility away from themselves~\cite{lima2022explainable}.

Our findings demonstrate how laypeople's moral judgments of machines and humans differ. Even though fairness is a well-established factor in moral reasoning about human actors, it has no effect on blame judgments of AI systems. Instead, judgments about AI systems are determined by people's stance towards the possibility of blaming machines. Those who believe AI systems are not suitable recipients of blame choose not to blame them, whereas those who believe otherwise blame machines as much as other actors. We discuss how future decisions to integrate AI systems into the moral and social spheres will affect how people react to their actions.

%% file: content/2background.tex
\label{sec:background}

\subsection{Blameworthy AI Systems?}

Automated systems, such as AI and robots, are becoming more common in high-risk scenarios. For instance, they assist doctors prioritize patients who require urgent medical care and aid employers review job applicants to expedite the hiring process. They help judges make bail decisions and operate self-driving vehicles. These systems, however, are far from perfect. Decision-making algorithms have been found to discriminate against African American patients and defendants~\cite{obermeyer2019dissecting,propublicastory}. There have been stories of AI recruiting tools disfavouring women in the hiring process~\cite{reutersamazon} and self-driving cars killing pedestrians and passengers~\cite{nytteslapedestrian,nytteslapassengers}. This study looks into how laypeople, i.e., those subjected to AI systems, react to the harms these new technologies cause.

An extreme example of public reaction to harm caused by automated systems is the vandalism against autonomous vehicles following the killing of a pedestrian by a self-driving car~\cite{nytknives}. According to~\citet{liu2021blame}, people overreact to crashes caused by autonomous vehicles, which could hinder their adoption. In contrast, other research suggests that people may underreact when algorithms make biased decisions compared to when a human is the decision-maker~\cite{bigman2020algorithmic}. These contrasting examples highlight the complexity of understanding laypeople's reactions to algorithmic harm. Understanding how the public might react in such scenarios can help prevent future conflicts between the design of AI systems, their regulation, and public opinion~\cite{bonnefon2020moral,awad2020crowdsourcing}.

We study laypeople's reactions through the lens of blame. What it means to blame someone has been a contentious topic among moral philosophers. Scanlon~\cite{scanlon2013interpreting} argues that judgments of blameworthiness are based on assessments of an actor's attitudes in relation to what is expected from them. In this view, to blame someone is to modify one's relationship with the blamee according to their perceived blameworthiness. Another perspective is put forward by Strawson~\cite{strawson20181}, who suggests that blaming someone consists of a series of emotional responses, i.e., the so-called ``reactive attitudes.'' Instead of defining blame based on the attitudes or actions of an actor, Shoemaker~\cite{shoemaker2021moral} defends that blaming functions to signal the blamer's commitment to a set of norms regardless of how it is operationalized.

Alongside the theoretical discussions around blame, empirical research suggests that people blame machines for the harm they cause. When robots do not make a utilitarian decision, they are blamed more than their human counterparts~\cite{malle2014theory}.\footnote{Although we also mention that \citet{lee2021people} provide a contrasting study, in which participants did not blame robots for their actions in the trolley dilemma.} Another study found that AI and human decision-makers are blamed similarly when it comes to bail decisions, despite the fact that automated systems were not given the same forward-looking responsibilities as their human counterparts~\cite{lima2021human}.\footnote{Backward-looking notions of responsibility (e.g., blame) pertain to previous actions, decisions, and consequences. Forward-looking responsibilities prescribe obligations concerning future actions~\cite{van2011relation}. Some suggest that not attending to an obligation (forward-looking responsibility) can lead to blame (backward-looking responsibility)~\cite{van2011relation}.} Autonomous systems are blamed more than those that rely on human input, implying that people's perceptions of autonomy play a key role in how they react to algorithmic  harm~\cite{kim2006should,furlough2021attributing}. Most research concludes that automated systems, such as AI systems, are more susceptible to blame than other machines (e.g., a vending machine) but not as much as humans~\cite{kahn2012people}.

Normative research has questioned the appropriateness and viability of individuals blaming automated systems. Scholars argue that blaming AI systems would be morally wrong because they are not appropriate subjects of retributive blame~\cite{danaher2016robots}. Machines lack the necessary moral understanding to be blamed~\cite{veliz2021moral} and do not possess the sentience required for comprehending what it means to harm someone~\cite{torrance2008ethics}. These arguments concern whether AI systems and robots have the properties necessary to be blamed~\cite{champagne2021mandatory}, and those who subscribe to this viewpoint agree that current automated systems do not.

In contrast, other scholars view blame as a social process that does not always track whether agents satisfy the necessary properties~\cite{tigard2021artificial}. Instead, they suggest people react to wrongdoers before considering their sentience, autonomy, or any other property; these considerations can only serve as a secondary (and possibly mitigating) factor. Proponents of this viewpoint argue that individuals can adapt their reactive attitudes to machines, particularly if they appear to be blameworthy~\cite{coeckelbergh2009virtual} or if doing so fulfills crucial social functions~\cite{stahl2006responsible}. This perspective appears to be consistent with cognitive science studies~\cite{malle2014theory,bigman2019holding} and the literature on how people blame AI systems and robots (see above).

It is unclear whether people apply similar moral frameworks when judging machines and humans. For instance, past research suggests that whether a human agent is supervised impacts how much they are blamed~\cite{hamilton1986chains,gibson2003ought}; similarly, machines perceived as highly autonomous are blamed more than their supervised counterparts ~\cite{kim2006should,furlough2021attributing,lima2021punish}. In contrast, it has also been shown that people may forgive humans but not machines for the same action~\cite{malle2019ai,scheutz2020may}, suggesting differences in how people react to harm depending on the actor.

Exploring whom people blame when machines cause harm is significant given the concern that the deployment of autonomous and self-learning systems gives rise to a responsibility gap~\cite{matthias2004responsibility}. This gap posits that autonomous and self-learning machines pose difficulties in holding human stakeholders responsible because of these systems' autonomy and adaptability. Blameworthiness is one of the components of the responsibility gap~\cite{de2021four}, raising the question of whether any actor is a suitable subject of blame when machines cause harm.

Although distinct from the question of who should answer legally and morally for machine-cause harms, i.e., who is accountable, deciding who is to blame can help determine ``contenders for the class of accountable actors,'' thus outlining who may face any potential sanctions for harmful outcomes~\cite{cooper2022accountability}. This relationship between blame and accountability can be exemplified by the concern that AI systems may become responsibility shields by absorbing blame that should have been attributed to human agents~\cite{bryson2017and,johnson2006computer}, exempting humans from the duty to answer for their actions and avoiding any penalties. This possibility highlights the importance of empirical research investigating how laypeople attribute blame to machines and human agents to ensure that accountability is not eroded when things go awry.

This research explores three factors that may influence the extent to which people blame AI systems and the actors developing and deploying them: 1) perceived harmfulness, 2) explainability, and 3) one's attitudes towards automated systems.

\subsection{Perceived Harmfulness}
Social psychology theories posit that perceived harm is the ``essence of morality''~\cite{gray2012mind}. Yet, research is yet to explore how the perceived harmfulness of algorithmic decisions affects to what extent AI systems are blamed. \citet{hidalgo2021humans} found that the same actions performed by humans and machines may be viewed as distinctively harmful. In this research, we explore how perceived harmfulness may impact blame judgments experimentally.

\subsection{Explainable AI}
Another factor we examine in this research is explainability. Most current AI systems are black-boxes whose outputs are often uninterpretable to human observers. As these systems start making consequential decisions, it becomes essential to understand how they work. This need has fueled the emergence of the field of explainable AI (XAI), which tries to create models ``that produce details or reasons to make [their] functioning clear or easy to understand''~\cite{arrieta2020explainable}. 

A review of the global guidelines for AI has found that explainability is the most prominent principle across all efforts to develop ethical AI systems~\cite{jobin2019global}. Explainable machines have a wide range of functions, ranging from ensuring that systems are fair to evaluating whether they do what they are supposed to~\cite{langer2021we}. We focus on the proposal of explainability to ensure that human actors can maintain meaningful human control over automated systems and thus remain responsible for any harm they may cause~\cite{robbins2019misdirected,langer2021we}. Explainable systems could fulfill legal requirements for holding developers and users responsible~\cite{bibal2020legal} and contribute to the perception that AI systems are \emph{not} to blame and that human actors are the ones that should be held responsible. Explainability should thus help maintain developers and users as appropriate subjects of responsibility~\cite{robbins2019misdirected,miller2019explanation,langer2021we}. An early study seems to point to the opposite direction: machines explaining why they committed a mistake decreased how much their users were blamed~\cite{kim2006should}.

\subsection{People's Attitudes Towards AI}
The last factor we investigate concerns people's perceptions of and attitudes towards AI systems. Scholars argue that the public perception of new technologies has crucial implications for how they will be adopted, developed, and regulated~\cite{cave2019hopes}. Both accurate and misleading narratives about AI can drive and hinder public acceptance, as shown in the past with genetically modified (GM) crops, whose reception was shaped by public risk perceptions~\cite{cave2018portrayals}. These attitudes can sway the design of machines and inform policymakers in both positive and negative directions~\cite{cave2019hopes}. Science fiction can also influence how people perceive these systems~\cite{cave2019hopes}; for instance, they may portray AI and robots as intelligent, sentient, and intentional, which should affect how people react to them. These portrayals could create false perceptions with moral and legal consequences to how laypeople interact with machines~\cite{bryson2010robots}.

Recent research in HCI provides some evidence that people may have different stances towards and expectations about human and AI actors. For instance, people exhibit varying expectations about human- and machine-led decisions in different types of decision-making tasks~\cite{langer2021look}. AI systems' decisions are perceived as less fair and trustworthy than identical human decisions in settings that require human skills (i.e., subjective tasks, such as hiring), but not in settings perceived as mechanical (i.e., objective tasks, such as work scheduling)~\cite{lee2018understanding,castelo2019task}. People exhibit a preference for humans over AI systems in high-stakes scenarios~\cite{langer2019highly} and in settings associated with higher degrees of inherent uncertainty~\cite{dietvorst2020people}, such as investing and medical decision-making. Past work indicates that making an algorithm transparent to users influences their views towards AI recommendations and decisions~\cite{eslami2019user,brown2019toward}, suggesting that the provenance of explanations can also impact lay perceptions of algorithmic decision-making (as explored in this research). Considering people's different expectations about humans and AI systems, we thus inquire whether lay perceptions of AI systems and their decisions influence how much they (and other actors) are blamed.

Lay perceptions of AI systems also differ based on one's understanding of AI. People have diverse mental models of AI systems, which are different across groups and over time, impacting how they perceive and use algorithms~\cite{eslami2015always,eslami2016first}. In this work, we do not attempt to create a standardized understanding of decision-making AI across participants (e.g., by defining it). Doing so would not reflect the real world~\cite{kapania2022because} but instead limit our efforts to capture and understand how people attribute blame when AI systems cause harm in the real world. In contrast, as shown below, we explore how these different lay attitudes towards AI systems (and the possibility of blaming them) help determine whether machines deserve blame.

\subsection{Research Preliminaries}
\label{sec:prelim}

In the present research, we focus on AI systems that make consequential decisions. Algorithms have been developed to assist humans making decisions in a wide range of scenarios (e.g.,~\cite{propublicastory,reutersamazon}), and regulatory proposals defend that AI systems should only be deployed with human oversight~\cite{veale2021demystifying}. Completely delegating decisions to algorithms may be illegal in certain domains, such as bail decisions, in which defendants have the right to be heard by a competent human judge~\cite{grgic2019human}. We note, however, that several AI systems currently take on the role of decision-makers in the real world. Existing algorithms decide who is automatically enrolled in health care programs~\cite{obermeyer2019dissecting} and determine which job postings are shown to online users, impacting their job prospects~\cite{datta2015automated}. Decision-making AI systems can also decide which images are returned by search results, which research has found to reinforce racial and gender stereotypes and shape people's views of their job opportunities~\cite{metaxa2021image}.

Even in scenarios where AI systems are deployed alongside humans, they may still be decision-makers in practice. Resume-screening algorithms decide which job applications should be evaluated by a human recruiter, rejecting an initial pool of applicants without human oversight~\cite{nytresume}. In scenarios where AI systems rank applicants instead of directly rejecting them, algorithms can still practically decide who will be evaluated by humans by determining who should appear on the first few result pages. Tenant-screening algorithms are marketed as decision-making products and disqualify applicants ''without providing any details [...] that would have allowed the property manager to make his own decision''~\cite{themarkuphousing}, obscuring human decision-making. Humans tend to overrely on machines~\cite{korber2018introduction} and decision-making algorithms~\cite{bansal2019beyond}, particularly if they are explainable~\cite{bansal2020does} as the ones explored in this research. Hence, AI systems can shape people's decision-making processes and limit their courses of action, taking on the role of a real decision-maker.

Finally, we clarify how we operationalize ``blame,'' which lacks a unified definition. We neither emphasize nor ask that our study participants adhere to a particular definition. Instead, we collect lay intuitions about blame, allowing participants to interpret what blame means and disclose their judgments accordingly. These judgments are referred to as \emph{``blame judgments''} in this work. Past research followed a similar approach to explore how people blame machines (e.g.,~\cite{malle2015sacrifice,malle2019ai}) and other human actors (see a review of the moral psychology literature on blame~\cite{malle2014theory}). Had we provided a particular definition of blame to participants, we would have biased their responses and measured how well they could follow the desired interpretation of blame. In contrast, our decision allowed us to explore how laypeople attribute blame to AI systems compared to human agents, expanding on the extensive literature investigating blame judgments of humans. As we will show below, lay judgments of AI systems have unique features that are not considered when blaming human agents, providing insights that could inform future research aiming to define what blame is and who (or what) can be blamed (see Section~\ref{sec:conclusion}).

%% file: content/3studyIntro.tex
\label{sec:studyIntro}

\begin{figure*}[!ht]
    \centering
    \includegraphics[width=0.9\textwidth]{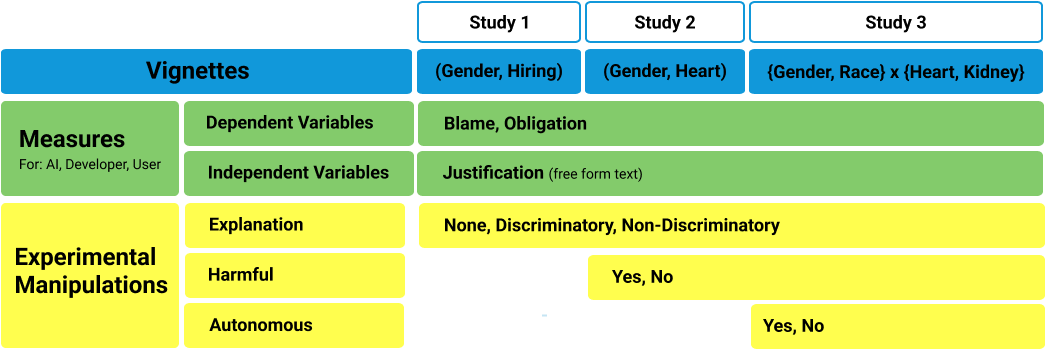}
    \caption{High-level overview of the methodology used in Studies 1, 2, and 3. The exact phrasing of the vignettes and questions can be found in the Appendix.}
    \Description{High-level overview of the methodology used in Studies 1, 2, and 3. Detailed methodology is presented in the main text and the Appendix.}
    \label{fig:methodology}
\end{figure*}

We present three studies exploring how the perceived explainability and harmfulness of algorithmic decisions, as well as people's attitudes towards AI systems, affect lay reactions to harms caused by machines. We formulate three research questions based on prior work presented in Section~\ref{sec:background}:

\begin{itemize}
    \item[RQ1)] How do post-hoc explanations impact blame judgments of AI systems, their developers, and users
    \item[RQ2)] How does perceived harmfulness of an algorithmic decision influence blame judgments of AI systems, their developers, and users?
    \item[RQ3)] How do one's attitudes towards AI systems---concerning their perceived autonomy---impact blame judgments of AI systems, their developers, and users?
\end{itemize}

Study 1 tested the effect of post-hoc explanations on blame judgments (RQ1) and identified two additional factors that correlated with how blame was distributed: perceived harmfulness and participants' attitudes towards AI systems. Study 2 investigated how perceived harmfulness influences people's judgments (RQ2), whereas Study 3 explored whether the perceived autonomy of AI systems is the main driver of the blame directed at them (RQ3). Study 3 was not successful in manipulating perceived autonomy; hence, we conducted a qualitative assessment of participants' open-ended justifications and found people's attitudes towards the possibility of blaming machines to be the most significant factor determining whether people blame AI systems. Figure~\ref{fig:methodology} presents a high-level overview of the three studies, their treatment conditions, and measures. All studies were approved by the first author's IRB, and we make our data and scripts available at \url{https://bit.ly/3DtFH7H}.

%% file: content/4study1.tex
The primary aim of Study 1 was to experimentally test the effect of explainability on how people blame AI systems alongside their developers and users (RQ1). This design was inspired by scholars suggesting that AI systems providing explanations for their decisions create the perception that human actors are the ones to blame. We also conducted an exploratory analysis investigating how perceived harmfulness correlates with laypeople's judgments and a qualitative analysis of people's open-ended explanations of their judgments.

\subsubsection{Study Design}

After agreeing to the research terms, participants were shown a three sentences-long introduction that explained how AI systems are currently used to make and assist decisions in a wide range of environments. Participants were then shown a vignette in which Systemy, a local technology firm, used an AI system to hire new software developers. The scenario presented Taylor, a junior software developer that applied for the position and was later rejected by the AI system. The AI system explained its decision on a between-subjects basis: it either did not give any explanation (\emph{no explanation/none}), explained that Taylor did not have the necessary experience (\emph{explanation}), or justified its decision based on gender (i.e., because Taylor is a woman; \emph{discriminatory explanation}).

A few design choices are worth clarifying. The hiring domain was selected due to the increasing use of decision-making systems in selection processes (e.g.,~\cite{nytresume}). We focused on gender-based discrimination, being inspired by recent incidents of hiring AI systems discriminating against women~\cite{reutersamazon}. The phrasing of the vignette was inspired by the work of~\citet{plane2017exploring}. We named the victim Taylor---a unisex name that is common among both White and African-Americans---to minimize the salience of the victim's gender and race in the treatments in which these features are not a part of the experimental manipulation.

Our treatment conditions were inspired by the natural language explanations employed by Facebook when justifying which advertisements are shown to its users~\cite{andreou2018investigating}. We employed simple text-based explanations to mitigate the effect of any confounding variables by presenting justifications that could be understood by any participant. The development of explainable algorithms aims to make them understandable to an audience, be it the general public, policymakers, or developers~\cite{arrieta2020explainable}. Hence, our study provided explanations that would be understandable to laypeople, i.e., the study's audience, by removing any complexity that could impact their interpretation. This design choice allowed us to measure the effect of explainability on blame judgments while mitigating the influence of latent variables, such as their ability to comprehend specific types of explanations.

Our study also investigated how discriminatory explanations impact blame judgments. We employed illegal explanations to ensure that our treatment condition was viewed as discriminatory.\footnote{Admitting that one's hiring decision is based on gender is illegal in the US and many other jurisdictions. That is not to say that gender-based discrimination in the job market does not exist; nevertheless, decision-makers would not give such an explicitly discriminatory explanation to avoid legal prosecution.} By presenting egregious justifications based on gender, we mitigated the effect of confounding variables that could have influenced people's judgments of fairness and blame. We acknowledge that most existing algorithms that were found to discriminate based on gender did so indirectly since they did not directly consider gender in their decision-making processes (e.g.,~\cite{datta2015automated,reutersamazon}). Hence, to ensure that our results could be extended to more realistic scenarios, we replicated our results with legal but clearly discriminatory explanations in Study 3.

\subsubsection{Measures}

Participants were first asked three questions: 1) how much \emph{blame} the AI system, its developer, and Systemy (i.e., its user) deserved for the decision not to hire Taylor and 2) how responsible these actors were for ensuring that the decision was correct, i.e., to what extent they prescribed an \emph{obligation} to each of the actors. Participants also 3) explained their judgments in a free text form. These three questions were shown separately for each of the actors. The presentation order of the actors was randomized between subjects.

Afterwards, respondents indicated to what extent the AI system explained its decision (\emph{perceived explanability}) and the extent to which the decision was fair (\emph{perceived fairness}) and harmful (\emph{perceived harmfulness}). All questions were answered using a 7-point scale coded from 0 to 6. We present all measures and materials in the Appendix. The survey ended with a series of demographic questions and debriefing.

\subsubsection{Participants} 

We conducted a power analysis to calculate the minimum sample size. A Wilcoxon-Mann-Whitney two-tailed test requires 67 respondents per treatment group to detect an effect size of Cohen's $d$ = 0.5 at the significance level of 0.05 with 0.8 power. Hence, we recruited 212 participants through the Prolific crowdsourcing platform~\cite{palan2018prolific} to account for attention-check failures, i.e., at least 67 participants per treatment group, assuming a 5\% attention-check failure rate. We targeted US residents that had previously completed at least 50 tasks on Prolific, with an approval rate of 95\% or above. Three participants failed at least one of the attention check questions that were presented before and after the vignette, resulting in a final sample size of 209 (47.85\% women; $M_{age}$ = 36.08, $SD_{age}$ = 12.19). All participants were compensated a median of US\$7.30 per hour.

\subsubsection{Results}

\begin{figure*}[t!]
    \centering
    \includegraphics[width=\textwidth]{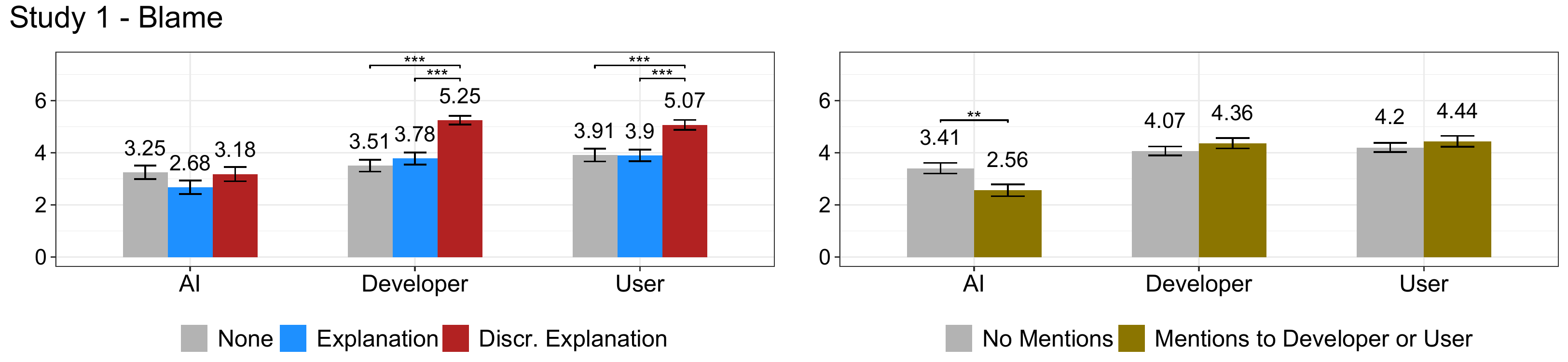}
    \caption{Blame judgments of the AI system, its developer, and user involved in making hiring decisions in Study 1. Participants were randomly assigned to a treatment condition where the AI system either did not provide any explanation for its decision (i.e., \emph{None}), justified its reasoning based on the applicant's experience (i.e., \emph{Explanation}), or discriminated against the applicant based on her gender (i.e., \emph{Discr. Explanation}). Participants were also categorized into those that explained their blame judgments of the AI system with \emph{mentions to its developer or user} and those who did not highlight the role of the human actors developing and deploying the AI system (i.e., \emph{no mentions}). Standard errors are presented as error bars. $^{*}$\pvalue{.05}, $^{**}$\pvalue{.01}, $^{***}$\pvalue{.001}.}
    \Description{Blame judgments of the AI system, its developer, and user involved in making hiring decisions in Study 1. Explainability alone did not influence blame directed at any of the actors. Discriminatory explanations increased blame towards the user and developer, but not the AI system. Participants that mentioned the developer or user in their blame judgments of the AI system blamed the machine less. Please refer to the main text for effect sizes and the Appendix for numerical values.}
    \label{fig:s1_blame}
\end{figure*}

We employed one-way analysis of variance (ANOVA) tests to identify differences between treatment conditions. Decisions followed by a discriminatory explanation were perceived as more unfair than those complemented by a justification based on the applicant's experience or no explanation at all (\ftest{2}{206}{50.15}, \pvalue{.001}, \etasq{0.33}). Explainable systems were also regarded as more explainable than their opaque counterparts regardless of whether their explanations were discriminatory or not (\ftest{2}{206}{24.89}, \pvalue{.001}, \etasq{0.19}). These results show that the between-subjects conditions achieved the desired effect on perceived explainability and fairness (see Appendix for mean values and post-hoc pairwise tests).

Figure~\ref{fig:s1_blame} presents participants' blame judgments of the AI system, its developer, and user (see Appendix for numerical values and Tukey's HSD post-hoc tests). Explainability did not influence the extent to which the AI system was blamed (\ftest{2}{206}{1.35}, \pvalueequal{.262}, \etasq{0.01}). 
In contrast, the developer of a discriminatory AI system was blamed more than one that developed a non-discriminatory explainable system or an opaque decision-maker (\ftest{2}{206}{21.01}, \pvalue{.001}, \etasq{0.17}). 
Similarly, participants blamed the user of a discriminatory AI system more than other users (\ftest{2}{206}{9.30}, \pvalue{.001}, \etasq{0.08}).

Participants' attribution of obligations to all actors was not influenced by the study treatment (\pvaluegreater{.05} for all actors). Nonetheless, the AI system (\msd{2.66}{2.16}) was given obligations to a lesser extent than its developer (\msd{4.10}{1.96}) or user (\msd{4.03}{2.07}).

An exploratory analysis suggested that perceived harmfulness correlated with blame judgments of all actors, such that participants who perceived the decision as more harmful blamed the AI system (\corr{0.20}, \pvalue{.005}), its developer (\corr{0.49}, \pvalue{.001}), and user (\corr{0.35}, \pvalue{.001}) more. In contrast, perceived harmfulness was not associated with attribution of obligations to any of the actors (all \pvaluegreater{.05}).

Investigating participants' open-ended justifications for their blame judgments of the AI system suggested that some rationalized their reactive attitudes based on the perception that these systems are programmed and used by humans, and thus should not be blamed. To identify those who mentioned the AI system's developer or user in their justifications, we used the regular expression \textit{``systemy|dev|progr|company|user|medical|hospi\-tal|west|doct''} to identify participants that highlighted the role of developers or users in their blame judgments of the AI system. This regular expression was crafted after a qualitative analysis of participants' open-ended responses and aimed to identify those who justified their blame judgments of machines with mentions to human and collective agents (e.g., the programmer, Systemy). This regular expression was iteratively updated throughout the three studies presented in this paper and contains terms used to identify the AI system's user in Studies 2 and 3. We categorized participants into two groups: those who mentioned the AI system's developer or user in their explanation (43.54\%) and those who did not (56.46\%).

We used a 3 (treatment) x 2 (mentions or not) ANOVA to account for this variable (see Figure~\ref{fig:s1_blame}). We found that participants who mentioned the developer or user blamed the AI system less than those who did not (\ftest{1}{203}{7.52}, \pvalue{.01}, \etasq{0.04}). In contrast, this variable did not influence blame judgments of developers (\ftest{1}{203}{0.32}, \pvalueequal{.574}, \etasq{0.00}) and users (\ftest{1}{203}{0.26}, \pvalueequal{.610}, \etasq{0.00}). The main effects of the explainability treatment were consistent with our initial analysis, and we did not observe any significant interaction between these two factors (all \pvaluegreater{.05})

\subsubsection{Discussion}

\h{Explainability, by itself, did not impact the extent to which the AI system, its developer, and user were blamed. However, the perceived fairness of the AI system's explanation did.} The developers and users of an AI system that provided a discriminatory explanation received more blame, while reactive attitudes directed at the machine did not change accordingly. Surprisingly, blame towards the AI system did not increase when it provided an egregious and unfair explanation based on gender. With respect to explainability, explainable AI systems were not blamed to a larger extent than their opaque counterparts. Their developers and users were also blamed similarly to those of a non-explainable system. This result contrasts with the proposal of explainability to facilitate identifying which human agent is responsible when machines cause harm.

An exploratory analysis of Study 1 suggested that perceived harmfulness might influence blame judgments of machines and other actors differently. However, our scenario was not perceived as particularly harmful; mean perceived harmfulness was close to the mid-point (\msd{3.86}{1.81}). A possible explanation is that the rejection of a job application might have been viewed as an opportunity loss rather than a clearly harmful outcome. Moreover, the design of Study 1 did not control for the interaction between perceived fairness and harmfulness, preventing us from distinguishing their distinct effects on blame judgments. We examine this factor further in Studies 2 and 3.

\h{Nevertheless, our results suggest that the main factor at play in blame judgments of machines may not be perceived harmfulness or fairness but people's perceptions of these systems.} Those who justified their blame judgments by mentioning the developer or user tended to blame the machine less than those who did not---a finding we explore further in subsequent studies and Section~\ref{subsec:qualitative}.

Finally, we observed that AI systems were attributed obligations (e.g., forward-looking responsibilities) to a lesser extent than their developers and users, although they did not differ between treatment conditions. These results agree with previous work, which has shown that machines are attributed less forward-looking notions of responsibility than humans~\cite{lima2021human} and that such judgments are indifferent to an action's consequences~\cite{van2011relation}.

%% file: content/5study2.tex
\begin{figure*}[!ht]
    \centering
    \includegraphics[width=\textwidth]{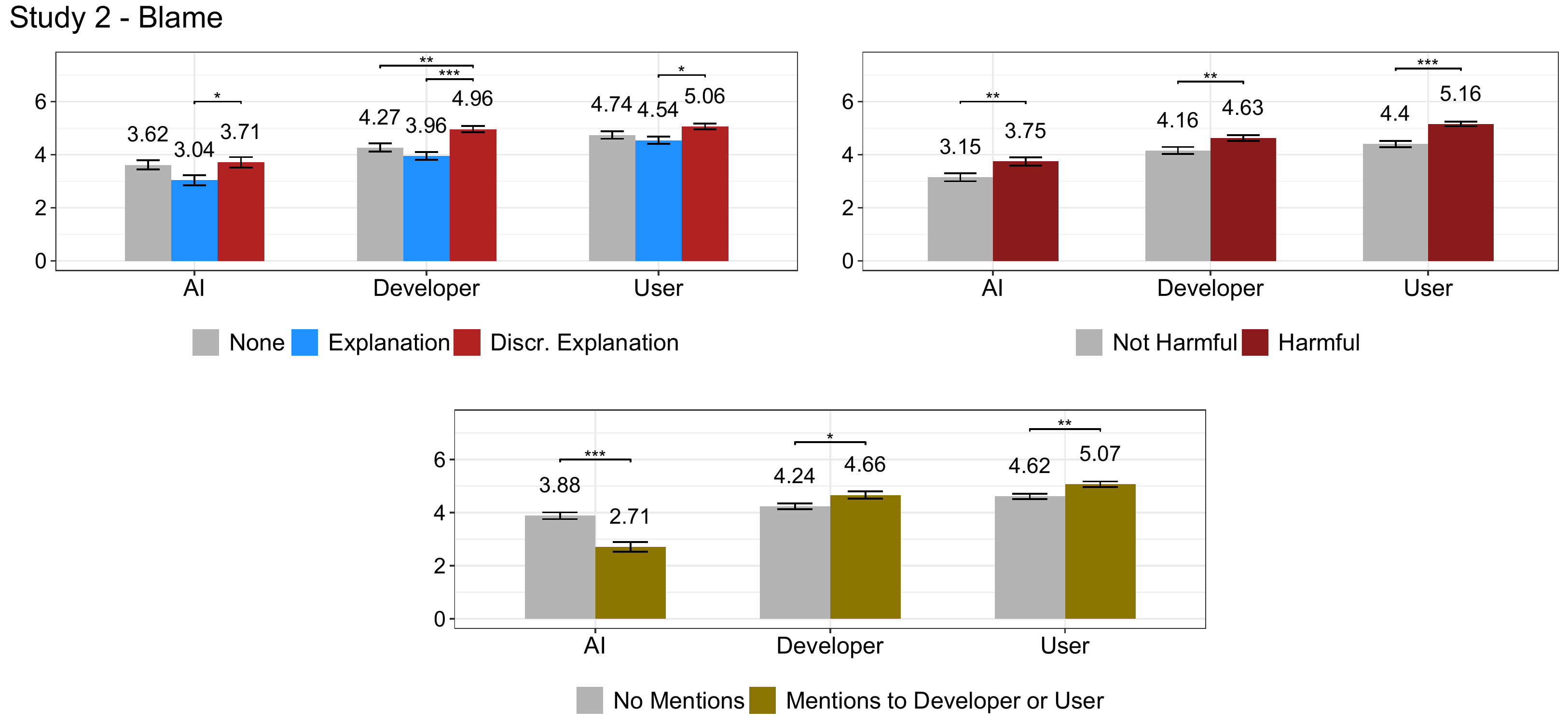}
    \caption{Blame judgments of the AI system, its developer, and user involved in making medical prioritization decisions in Study 2. Participants were randomly assigned to a treatment condition where the AI system either did not provide any explanation for its decision (i.e., \emph{None}), justified its reasoning based on the patient's medical pre-conditions (i.e., \emph{Explanation}), or discriminated against the patient based on her gender (i.e., \emph{Discr. Explanation}). The vignette concluded with either the patient receiving medication for her symptoms (i.e., \emph{Not Harmful}) or dying without receiving proper medical care (i.e., \emph{Harmful}). Participants were also categorized into those that explained their blame judgments of the AI system with \emph{mentions to its developer or user} and those who did not highlight the role of the human actors developing and deploying the AI system (i.e., \emph{no mentions}). Standard errors are presented as error bars. $^{*}$\pvalue{.05}, $^{**}$\pvalue{.01}, $^{***}$\pvalue{.001}.}
    \Description{Blame judgments of the AI system, its developer, and user involved in making medical prioritization decisions in Study 2. Explainability alone did not significantly influence blame directed at any of the actors. Discriminatory explanations increased blame towards the user and developer but not the AI system. All actors were blamed more when the decision led to the death of the patient. Participants that mentioned the developer or user in their blame judgments of the AI system blamed the machine less, while blaming the developer and user marginally more. Please refer to the main text for effect sizes and the Appendix for numerical values.}
    \label{fig:s2_blame}
\end{figure*}

Study 2 investigated how perceived harmfulness influences blame judgments of machines and the human actors involved in algorithmic decision-making experimentally (RQ2). Judgments of harmfulness in Study 1 were associated with perceived fairness; Study 2 detached these two factors and inquired how they may add up or interact in how people react to harm caused by AI systems.

We also complemented our research with a different domain: medical decision-making. Medicine has received considerable attention from AI research, with algorithms being used to discover new drugs~\cite{fleming2018artificial}, identify patients at risk~\cite{obermeyer2019dissecting}, and many other functions. We employed a vignette depicting disparities in the medical treatment of men and women~\cite{institute2010women}, specifically in terms of the diagnosis and treatment of heart attacks.
Coronary heart disease is underdiagnosed and undertreated in women~\cite{mehta2016acute}, and women are less likely to survive an acute myocardial infarction (commonly known as a heart attack), partially due to the differences in the clinical presentation~\cite{zucker1997presentations} and the medical treatment they receive~\cite{clarke1994women}. 
In contrast to the vignette employed in Study 1, the medical scenario allowed us to vary perceived harmfulness experimentally.

\subsubsection{Study Design}

Participants took part in a study similar to Study 1. Study 2 introduced West Medical, a local hospital that uses an AI system to assess the condition of emergency room patients. The scenario then presented Taylor, a woman that went to West Medical's emergency room after feeling discomfort in her left arm and nausea for three days straight, i.e., common symptoms of a heart attack. Taylor was classified as a non-critical patient by the AI system and either received no justification (\emph{no explanation/none}), was told that she was classified as such because she did not have any pre-existing conditions (\emph{explanation}), or was informed that this classification was due to her gender (i.e., because Taylor is a woman; \emph{discriminatory explanation}). Similarly to Study 1, we employed egregious explanations based on gender to mitigate the effect of confounding variables in the discriminatory explanation condition.

Study 2 also varied the outcome of the AI system's decision. After waiting for eight hours, Taylor was either examined by a doctor and prescribed some medication (\emph{not harmful}) or had a heart attack and died without receiving proper medical care (\emph{harmful}). All treatment conditions were randomly assigned between-subjects, such that each respondent read one of the 3 (explainability treatment) x 2 (harmfulness treatment) vignettes.

\subsubsection{Measures}

Participants answered the same set of questions from Study 1. All questions were answered using a 7-point scale coded from 0 to 6. See Appendix for details.

\subsubsection{Participants}

Considering the same power analysis from Study 1, which required 67 respondents per treatment group, we recruited 410 participants through Prolific. None of the participants had participated in Study 1. We employed the same recruitment and exclusion criteria as in Study 1. Four participants failed the attention check, resulting in a final sample size of 406 (46.55\% women; $M_{age}$ = 34.37, $SD_{age}$ = 11.81). All participants were compensated a median of US\$7.66 per hour.

\subsubsection{Results}

We employed 3 (explainability treatment) x 2 (harmfulness treatment) ANOVA models. The Appendix contains an analysis of perceived explainability, fairness, and harmfulness as manipulation checks. Our experimental manipulations were effective: explainable machines were perceived as more explainable, discriminatory decisions were viewed as more unfair, and harmful consequences were judged as more harmful.

Figure~\ref{fig:s2_blame} presents blame judgments of the AI system, its developer, and user by treatment condition in Study 2. The interaction terms between the two treatment conditions were not significant (\pvaluegreater{.05}) in blame judgments of all actors. Hence, we present mean values and standard errors independent of any interaction effects. The Appendix contains descriptive statistics and Tukey's HSD test results. 

Explainability had a small effect on judgments of the AI system: a machine that explained its decision without explicit discrimination was blamed marginally less than its discriminatory counterpart (\ftest{2}{400}{3.88}, \pvalue{.05}, \etasq{0.02}). An AI system whose decision led to the death of the patient was blamed more (\ftest{1}{400}{8.54}, \pvalue{.005}, \etasq{0.02}).

Blame directed at the AI system's developer was highest for that of a discriminatory machine (\ftest{2}{400}{13.31}, \pvalue{.001}, \etasq{0.06}). The effect of harmfulness was also significant: the developer of a system that led to the death of the patient received more blame (\ftest{1}{400}{9.25}, \pvalue{.005}, \etasq{0.02}). 

Harmfulness was a more significant factor in blame judgments of the AI system's user (i.e., West Medical; \ftest{1}{400}{27.98}, \pvalue{.001}, \etasq{0.07}). The user of a machine whose decision led to the patient's death was blamed more. Explainability influenced reactive attitudes towards the user to a lesser degree (\ftest{2}{400}{4.24}, \pvalue{.05}, \etasq{0.02}), such that users of a discriminatory system received more blame than those of a non-discriminatory explainable algorithm.

None of the treatment conditions influenced participants' attribution of obligations to the AI system, its developer, and user (\pvaluegreater{.05} for all main effects). This finding is consistent with Study 1. The only significant yet small effect was observed in the interaction between explainability and harmfulness in attributions of obligations to the user (\ftest{2}{400}{3.97}, \pvalue{.05}, \etasq{0.02}). Overall, the AI system (\msd{3.26}{2.26}) was ascribed forward-looking responsibilities to a lesser extent than its developer (\msd{4.40}{1.78}) and user (\msd{4.71}{1.70}).

Similarly to Study 1, we examined participants' justifications for their blame judgments of the AI system. Using the same categorization method as in Study 1, 36.70\% of participants mentioned the developer or user when explaining their judgments. We employed 3 (explainability) x 2 (harmfulness) x 2 (mentions to developer or user or not) ANOVA models to identify the effect of this variable on how people blame all actors. The AI system was blamed significantly more when participants did \emph{not} mention its developer or user (\ftest{1}{394}{32.77}, \pvalue{.001}, \etasq{0.08}). In contrast, the AI system's developer received more blame when participants mentioned the developer or user (\ftest{1}{394}{4.44}, \pvalue{.05}, \etasq{0.01}). The user was also blamed more when mentioned (\ftest{1}{394}{6.85}, \pvalue{.01}, \etasq{0.02}). The effect of explainability and harmfulness on judgments of all actors did not change from our initial analyses.

\subsubsection{Discussion}

\h{Both explainability and harmfulness influenced how much participants blamed the AI system's developer and user. As in Study 1, however, it was not explainability itself that increased blame directed at these entities; instead, it was the perceived fairness of the AI system's explanation. Importantly, the effect size of each treatment condition differed by actor.} Blame directed at the AI system's developer was influenced the most by fairness considerations (\etasq{0.06}, i.e., a medium effect size). In contrast, reactive attitudes towards the user were mostly determined by the perceived harmfulness of the decision (\etasq{0.07}). 
These results suggest that blame could be partly determined by the level of control actors hold over the decision-making algorithm. Developers can control which explanations are implemented in the AI system, whereas users can only decide whether to adopt the system and how it may impact those subjected to it.

Both explainability and harmfulness showed only a small effect on how participants blamed the AI system (\etasq{0.02}, i.e., a small effect size). Once more, the AI system was not blamed to a larger degree after justifying its decision with a clearly illegal and unfair explanation. As in Study 1, participants seem to judge machines differently from other actors. \h{Reactive attitudes towards the AI system were associated with how participants justified their judgments (\etasq{0.07}); the AI system was blamed depending on the perceived roles of its developer and user.} Investigating participants' open-ended responses suggested that participants who view the AI system as constrained by its programming blame it less, a hypothesis we explored in Study 3.

%% file: content/6study3.tex
\begin{figure*}[t!]
    \centering
    \includegraphics[width=\textwidth]{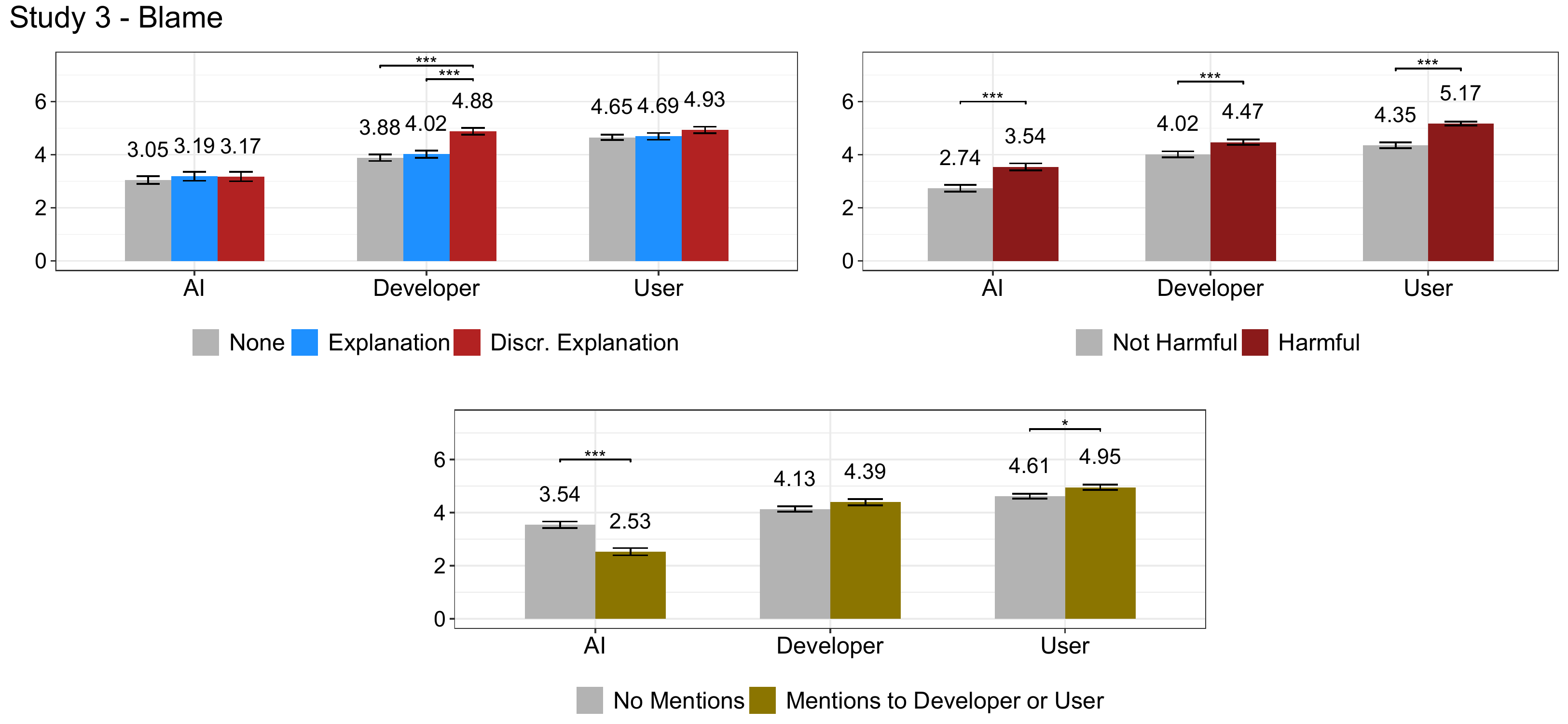}
    \caption{Blame judgments of the AI system, its developer, and user involved in making medical prioritization decisions in Study 3. Participants were randomly assigned to a treatment condition where the AI system either did not provide any explanation for its decision (i.e., \emph{None}), justified its reasoning based on the patient's medical pre-conditions (i.e., \emph{Explanation}), or discriminated against the patient based on their race or gender (i.e., \emph{Discr. Explanation}). The vignette concluded with either the patient receiving medication for their symptoms (i.e., \emph{Not Harmful}) or dying without receiving proper medical care (i.e., \emph{Harmful}). Participants were also categorized into those that explained their blame judgments of the AI system with \emph{mentions to its developer or user} and those who did not highlight the role of the human actors developing and deploying the AI system (i.e., \emph{no mentions}). Standard errors are presented as error bars. $^{*}$\pvalue{.05}, $^{**}$\pvalue{.01}, $^{***}$\pvalue{.001}.}
    \Description{Blame judgments of the AI system, its developer, and user involved in making medical prioritization decisions in Study 3. Explainability alone did not significantly influence blame directed at any of the actors. Discriminatory explanations increased blame towards the user and developer but not the AI system. All actors were blamed more when the decision led to the death of the patient. Participants that mentioned the developer or user in their blame judgments of the AI system blamed the machine less, while blaming the user marginally more. Please refer to the main text for effect sizes and the Appendix for numerical values.}
    \label{fig:s3_blame}
\end{figure*}

Study 3 had two major objectives. The first was to examine the effect of the AI system's perceived autonomy on how people react to algorithmic harm (RQ3). We test the hypothesis that people's judgments of AI systems are influenced by the perception that these machines' behavior is restricted by their programming, as suggested by Study 2 and prior work~\cite{furlough2021attributing,kim2006should}. The second objective was to replicate our findings using different vignettes in the medical domain. The discriminatory explanations presented in Studies 1 and 2 were not realistic as they justify the decisions by describing procedures that are largely illegal in the real world---namely, direct discrimination based on legally protected features. Hence, Study 3 additionally employed discriminatory explanations that could be reasonably adopted in the medical domain.

\subsubsection{Study Design}

Participants took part in a study similar to Study 2 but with two major modifications. First, participants were randomly assigned to a vignette that varied with respect to both the characteristics of the patient looking for medical assistance and the condition from which the patient was suffering. We employed 2 x 2 vignettes that introduced Taylor, who was either a woman or an African American person, suffering from symptoms suggesting either a heart attack or kidney failure. In addition to gender-based discrimination, Study 3 considered racial discrimination in the medical domain~\cite{ahmed2007racial}. We added vignettes related to kidney failure, inspired by research that has found that African American people are less likely to receive optimal treatment for chronic kidney disease and are more likely to progress to kidney failure~\cite{norris2021removal}. In contrast to the explanations based on gender used in Studies 1 and 2, which would not be used in practice, race may indeed be used as a variable in medical decisions related to kidney disease~\cite{ahmed2021examining}. The vignettes also included our prior explainability and harmfulness treatments, which were modified accordingly. For instance, Taylor was discriminated against based on race if introduced as African American. Participants were randomly assigned to one of the four vignettes.

Second, we included an additional treatment condition. After introducing West Medical and its usage of an AI system for patient prioritization, some participants were told that the AI system learned which patients to prioritize without explicit rules set by its developer (\emph{autonomous}). Other participants were not shown any information about how the AI system learned how to make decisions (\emph{N/A}); this treatment sought to understand how people's conception that these systems are under the control of their developers influences how machines are blamed. Hence, participants were randomly assigned between-subjects to one of 3 (explainability) x 2 (harmfulness) x 2 (autonomy) treatment conditions.

\subsubsection{Measures}

Participants answered the same set of questions from Study 2. In addition, participants indicated the extent to which they thought ``the AI system's decision was under the control of its developer,'' ``the AI system was programmed to make this decision,'' and ``the AI system's decision was explicitly programmed by its developer.'' These questions were used as manipulation checks for our autonomy treatment condition. All questions were answered using a 7-point scale coded from 0 to 6. The Appendix presents all measures and materials.

\subsubsection{Participants}

We recruited 550 participants via Prolific using the same recruitment and exclusion criteria as earlier, out of whom 12 failed the attention check. No participant had participated in the previous studies. Our final sample comprised 538 participants (45.91\% women; $M_{age}$ = 35.53, $SD_{age}$ = 11.71), achieving a power of 0.85 to detect the smallest effect size we had found in previous studies (\etasq{0.02}) at a significance level of 0.05. All participants were compensated at a median rate of US\$10.35 per hour.

\subsubsection{Results}

We employed 3 (explainability) x 2 (harmfulness) x 2 (autonomy) ANOVA models and present a comprehensive manipulation check analysis in the Appendix. Our explainability, fairness, and harmfulness manipulations obtained the desired effect. Perceived fairness did not differ between the illegal scenarios, in which Taylor was discriminated against based on gender, and the lawful vignettes, in which race was given as an explanation (\ftest{3}{523}{1.56}, \pvalueequal{.199}, \etasq{0.01}). However, the newly proposed autonomy treatment did not obtain the desired effect; the condition did not significantly increase perceived autonomy (average of responses to the three autonomy-related questions, Cronbach's $\alpha$ = 0.82; \ftest{1}{523}{6.00}, \pvalue{.05}, \etasq{0.01}). An initial analysis of blame judgments did not suggest a significant main effect of the autonomy manipulation factor (all \pvaluegreater{.05}).

Hence, we discarded the autonomy treatment for all subsequent analyses as it did not achieve the desired effect. Adding it to our analysis did not modify our results. We present findings from 3 (explainability) x 2 (harmfulness) x 2 (mentions to developer or user---40.71\% of participants---or not) ANOVA models, as done in Study 2. Analyses of participants' judgments using a 2 (explainability) x 2 (harmfulness) ANOVA model were indistinguishable from the reported results. We also include the vignette shown to participants as a fixed effect in all models.

Figure~\ref{fig:s3_blame} presents participants' blame judgments in Study 3 (see Appendix for descriptive statistics and Tukey's HSD post-hoc tests). Overall, our results are consistent with the findings from Studies 1 and 2. Some of the interaction terms between the treatment conditions were only marginally significant ($\eta_{p}^{2}$ $\le$ .01). We present mean values independently of these interactions. 

Participants that justified their judgments of the AI system with mentions to its developer or user blamed the machine less than those who did not (\ftest{1}{523}{31.28}, \pvalue{.001}, \etasq{0.06}). An AI system whose decision led to the death of a patient was blamed more than its counterpart (\ftest{1}{523}{21.00}, \pvalue{.001}, \etasq{0.04}). Explainability did not influence people's reactive attitudes towards the AI system (\ftest{2}{523}{0.82}, \pvalueequal{.443}, \etasq{0.00}).

The developer of a discriminatory AI system was blamed more than that of a non-discriminatory explainable machine or opaque system (\ftest{2}{523}{20.45}, \pvalue{.001}, \etasq{0.07}). Harmfulness also influenced the extent to which the AI system's developer was blamed (\ftest{1}{523}{13.64}, \pvalue{.001}, \etasq{0.03}). The developer of a machine that led to the death of the patient received more blame. Whether participants mentioned the AI system's developer or user when justifying their judgments of machines did not impact participants' reactive attitudes towards the developer (\ftest{1}{523}{1.80}, \pvalueequal{.180}, \etasq{0.00}).

The main factor at play in judgments of the AI system's user was the decision's perceived harmfulness (\ftest{1}{523}{43.46}, \pvalue{.001}, \etasq{0.08}). Users of a system that led to the death of the patient were blamed more than their counterparts. We also observed a small effect of people's justification of their blame judgments of machines (\ftest{1}{523}{5.58}, \pvalue{.05}, \etasq{0.01}), such that those who mentioned the developer or user blamed the user more. Explainability had a marginal effect on people's judgments (\ftest{2}{523}{3.28}, \pvalue{.05}, \etasq{0.01}), as did the vignette (\ftest{3}{523}{2.99}, \pvalue{.05}, \etasq{0.02}).

We observed a series of small effects in people's attribution of obligations to the AI system, its developer, and user (all $\eta_{p}^{2}$ \textless~0.03). Overall, the AI system was deemed less responsible in a forward-looking manner (\msd{3.27}{2.10}) than its developer (\msd{4.33}{1.72}) and user (\msd{4.93}{1.51}).

\subsubsection{Discussion}

Study 3 replicated our previous findings concerning the effect of explainability, fairness, and harmfulness on blame judgments of AI systems, their designers, and users in a wider range of vignettes. As in previous studies, explainability alone did not increase blame directed at any of the entities, which is in contrast to the proposal of explainability to facilitate identifying responsible human actors. \h{Nonetheless, our results demonstrate how different moral considerations shape blame directed at developers and users distinctively. Fairness judgments are more strongly associated with blame towards the AI system's developer, whereas judgments of users are more influenced by perceived harmfulness.}

Perceived fairness did not influence reactive attitudes towards AI systems. These results were found for the egregious explanations employed in Studies 1-3 and the lawful yet discriminatory justifications based on race from Study 3. Although perceived harmfulness seems to influence blame towards AI systems, its effect is small and dominated by people's perceptions of these systems.

From Study 2's findings, we hypothesized that people's belief that the programming of AI systems constrains their behavior would be the main determinant of their blame judgments. We thus attempted to experimentally manipulate the perceived autonomy of the AI system. However, the treatment did not significantly impact people's perceptions, suggesting that these beliefs may not be easily mutable. This negative result does not imply that perceived autonomy does not impact blame judgments, which would challenge prior work~\cite{kim2006should,furlough2021attributing}. To better understand what specific attitudes may influence blame towards AI systems, we expand our analysis of how people justify their judgments of machines in Section~\ref{subsec:qualitative}.

%% file: content/7meta.tex
\begin{figure*}[t!]
    \centering
    \includegraphics[width=\textwidth]{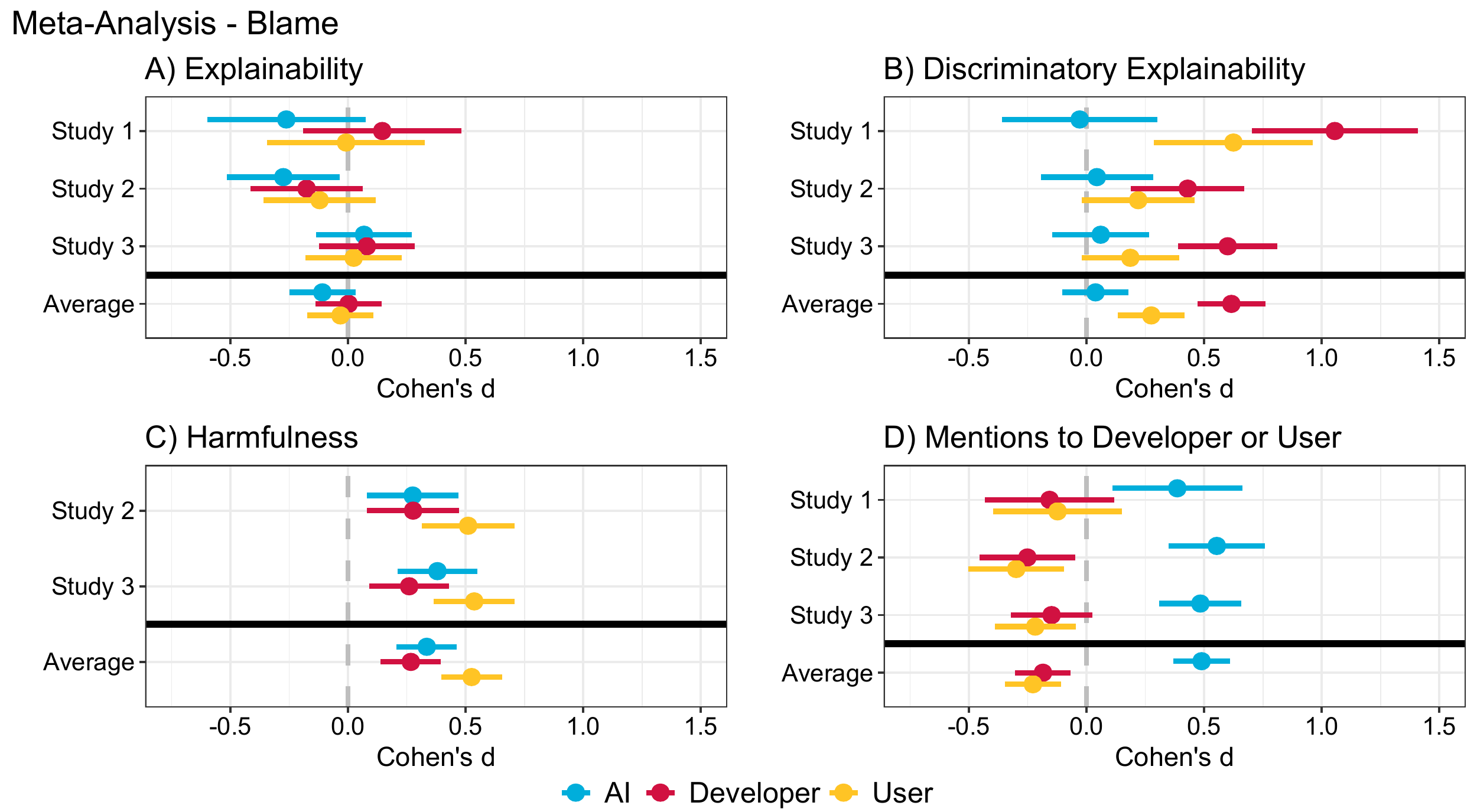}
    \caption{Meta-analysis of blame judgments of the AI system, its designer, and user. We present effect sizes (Cohen's $d$) between the treatment and control conditions across all studies and variables. We used a fixed-effects meta-analysis model and present 95\% confidence intervals as error bars.}
    \Description{Meta-analysis of blame judgments of the AI system, its designer, and user. We present effect sizes (Cohen's $d$) between the treatment and control conditions across all studies and variables. We used a fixed-effects meta-analysis model. Explainability did not significantly impact blame towards any of the actors. Discriminatory explanations increased blame towards the developer and user, and its effect size is significantly higher for the developer. Perceived harmfulness was associated with blame judgments of all actors, increasing blame for the user more than for others. Participants who mentioned the developer or user in their blame judgments of machines blamed the AI system more, while decreasing their reactive attitudes towards human actors.}
    \label{fig:meta}
\end{figure*}

Following recent best-practices recommendations~\cite{mcshane2017single}, we present a within-paper meta-analysis to explore the average effect size of our treatment conditions across the three studies. Researchers caution against relying on single studies when evaluating the robustness and reliability of an effect, and an internal meta-analysis can help synthesize findings from multiple studies~\cite{goh2016mini}. Hence, we conducted a meta-analysis of blame judgments using the \textit{metacont} function from the \textit{meta} package for the R programming language~\cite{schwarzer2007meta}. The meta-analysis calculates the treatments' weighted average effect sizes (Cohen's $d$) across all studies, which we present in Figure~\ref{fig:meta} alongside each study's estimates.

Whether the AI system was explainable did not impact judgments of any of the actors (Figure~\ref{fig:meta}A; RQ1); in contrast, discriminatory explanations influenced how much they were blamed (Figure~\ref{fig:meta}B; RQ1). This consideration had the largest effect on blame judgments of the AI system's developer and a significant but smaller effect on reactive attitudes towards its user. Perceived fairness did not impact judgments of the AI system; its effect size is close to zero across all studies.

Harmfulness influenced judgments of actors at different levels (Figure~\ref{fig:meta}C; RQ2). It had the largest effect on the blame directed at the AI system's user, while its effect on judgments of the machine and its developer was smaller. 
Finally, the main factor at play in how people blame the AI system can be captured by people's explanations of their judgments (Figure~\ref{fig:meta}D; RQ3). Those who did \emph{not} mention the developer or user when blaming the AI system directed blame towards the machine, mitigating their reactive attitudes towards human stakeholders. This finding led to the hypothesis that those who perceive the AI system's behavior as constrained by its programming blame the machine less. However, Study 3 was not successful in manipulating perceived autonomy of the system.
Considering the large effect sizes of people's justifications presented in Figure~\ref{fig:meta}D, we used a qualitative assessment of participants' explanations to understand which specific factors are associated with how people blame AI systems.

\section{Qualitative Analysis of Participants' Justifications}
\label{subsec:qualitative}

\begin{figure*}[t]
    \centering
    \includegraphics[width=.95\textwidth]{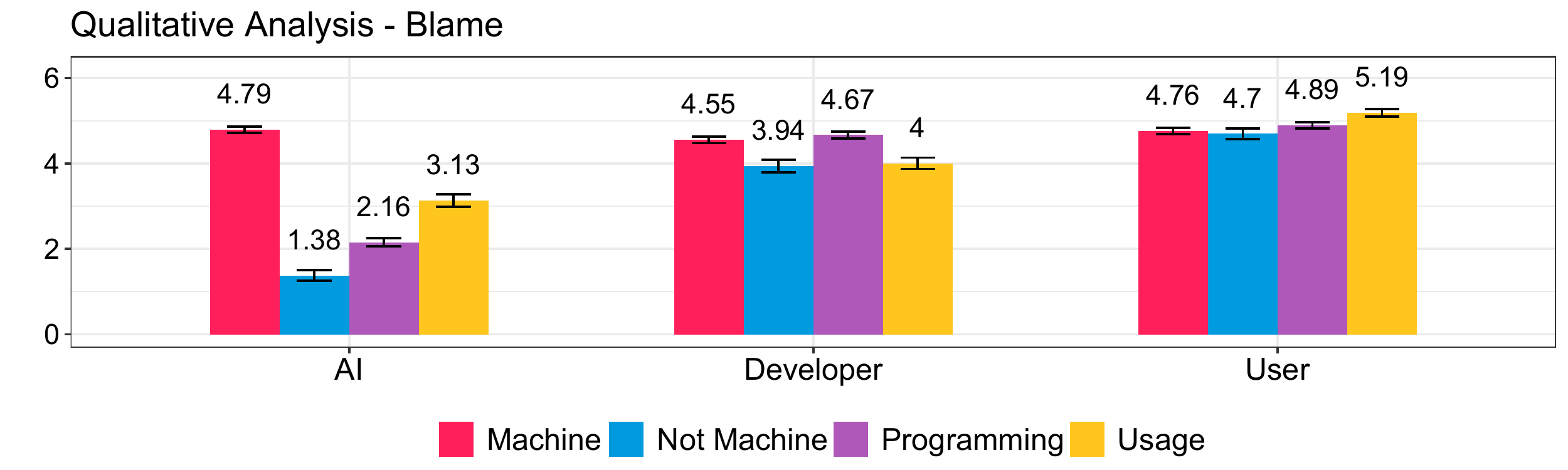}
    \caption{Blame judgments of the AI system, its developer, and user across all studies depending on participants' justifications of their blame judgments of the AI system. Participants' explanations were qualitatively coded into four categories (see Section~\ref{subsec:qualitative}). Standard errors are presented as error bars.}
    \Description{Blame judgments of the AI system, its developer, and user across all studies depending on participants' justifications of their blame judgments of the AI system. Participants' explanations were qualitatively coded into four categories. Those who view blaming a machine favorable blamed the AI system as much as other human actors, whereas those who denounce the possibility chose almost not to blame it. The perceived roles of developers and users influenced blame judgments of the respective actors and mitigated some blame towards the AI systems. Refer to the main text for the effect of each variable in detail.}
    \label{fig:qual}
\end{figure*}

We inspected participants' explanations manually; two authors grouped each response into several categories. We focused on people's justifications for their blame judgments of AI systems. The procedure was divided into three main steps. First, the authors coded a random sample of 5\% of responses to identify a series of categories; this initial procedure achieved high inter-coder reliability (Krippendorff's $\alpha$ = 0.879). Second, the authors coded an additional random sample of 10\% of responses to inspect whether the codes were sufficient and complete. There were no changes to the code list, and this step achieved high reliability ($\alpha$ = 0.868). Finally, the remaining responses were divided in half and coded separately. Disagreements and uncertain cases were discussed after independent coding for agreement. Responses were allowed to be coded as more than one category. The coding procedure identified four categories:
\begin{itemize}
    \item \underline{\textsf{Machine}}: Responses that justified the blame directed at the AI system by highlighting the machine's role in the decision-making process (e.g., ``the AI made the judgment'') or by explicitly stating that ``the system is to blame.''
    
    \item \underline{\textsf{Not Machine}}: Responses that denounced the idea of blaming AI systems, e.g., ``blame cannot be assigned to software'' or ``it cannot be held responsible.''
    
    \item \underline{\textsf{Programming}}: Responses that underscored the role that developers play in the process (e.g., ``AI was developed by humans'') or highlighted that the AI system's behavior is constrained by its programming (e.g., ``the system was just acting based on how it was programmed''). These participants emphasized some aspects of the AI system's programming.
    
    \item \underline{\textsf{Usage}}: Responses highlighting that there should be someone overseeing the AI system (e.g., ``humans should still be reviewing the decisions'') or stating that the user is to blame (e.g., ``the operator is responsible''). Overall, these responses focused on aspects of the AI system's usage.
\end{itemize}

Incomprehensible responses and participants that did not blame anyone because they deemed the decision correct (e.g., ``this was the correct course of action'') or did not blame a specific agent (e.g., ``too long wait'')---a total of 160 participants (13.87\%)---were discarded from this analysis.

\begin{table}[!t] \centering 
\small
\begin{tabular}{@{\extracolsep{5pt}}lccc} 
\toprule   & \multicolumn{1}{c}{AI} & \multicolumn{1}{c}{Developer} & \multicolumn{1}{c}{User} \\ 
& \multicolumn{1}{c}{(1)} & \multicolumn{1}{c}{(2)} & \multicolumn{1}{c}{(3)}\\ 
\midrule
Machine ($N$ = 417) & 1.765$^{***}$ & 0.328$^{*}$ & $-$0.040 \\ 
  & (0.158) & (0.154) & (0.140) \\ 
Not Machine ($N$ = 174) & $-$1.364$^{***}$ & $-$0.292$^{\dagger}$ & 0.021 \\ 
  & (0.168) & (0.163) & (0.149) \\ 
Programming ($N$ = 443) & $-$1.019$^{***}$ & 0.514$^{***}$ & 0.125 \\ 
  & (0.141) & (0.137) & (0.125) \\ 
Usage ($N$ = 194) & 0.098 & $-$0.364$^{*}$ & 0.446$^{***}$ \\ 
  & (0.149) & (0.145) & (0.132) \\ 
Intercept & 3.135$^{***}$ & 3.773$^{***}$ & 4.088$^{***}$ \\ 
  & (0.219) & (0.213) & (0.194) \\ 
\midrule
Adjusted $R^2$ & 0.438 & 0.105 & 0.077 \\
Observations & \multicolumn{1}{c}{993} & \multicolumn{1}{c}{993} & \multicolumn{1}{c}{993} \\ [1.8ex] 
\end{tabular} 
  \caption{Regression analysis of blame judgments of the AI system, its developer, and user as a function of participants' justification of their blame judgments towards the AI system. We report the number of responses coded as each category in the first column. Standard errors are shown inside parentheses. We present the complete results in Table~\ref{tab:full_regression} in the Appendix. $^{\dagger}$\pvalue{.1}, $^{*}$\pvalue{.05}, $^{**}$\pvalue{.01}, $^{***}$\pvalue{.001}.} 
  \label{tab:regression} 
\end{table}

We regressed participants' blame judgments of the AI system, its developer, and user in their 7-point scale to four dummy variables that were one based on the qualitative coding of the participants' justifications (see Table~\ref{tab:regression}). We also included dummy variables for the explainability and harmfulness treatments (and their interaction) to control for their effect (see Table~\ref{tab:full_regression} in the Appendix for the complete model). The study was incorporated as a fixed effect.

Figure~\ref{fig:qual} presents mean values of the responses coded as each category. People's stance towards AI systems and the possibility of blaming them played a significant role in how they reacted to algorithmic harm. Those who believe one can blame machines (coded as \underline{\textsf{Machine}}) directed their reactive attitudes at the AI system at a similar level to its designer and user. In contrast, a smaller but significant number of participants (\underline{\textsf{Not Machine}}) denounced this idea and chose to attribute almost no blame to the machine. Blame towards the AI system was also associated with how people viewed other actors' roles in the decision-making process. Interestingly, harmfulness considerations, which were significant in our previous analyses, did not impact judgments of the AI system anymore (see Table~\ref{tab:full_regression} in the Appendix). We see that the $R^2$ of the model predicting blame towards AI is much higher than those of the regressions modeling blame judgments of developers and users, highlighting how people's stance towards blaming machines is indeed the most significant factor in whether and how people blame AI systems. Removing these factors from the regression decreases the $R^2$ coefficient to 0.032.

Judgments of users were as expected; those who highlight the role of users in algorithmic decision-making (coded as \underline{\textsf{Usage}}) blamed them more. Similarly, blame judgments of AI developers were influenced by developers' perceived control over an AI system's decision (\underline{\textsf{Programming}}). Fairness and harmfulness considerations were still significant predictors of blame for both actors.

However, we also observed that those who highlight the AI system's agency (coded as \underline{\textsf{Machine}}) blamed developers more, whereas participants who denounce reactive attitudes towards machines (\underline{\textsf{Not Machine}}) tended to blame developers less. Although to a small degree, reactive attitudes towards the developer seem to track moral judgments of the AI system. It may well be that some participants equate the AI system and its developer (and not the user) when making moral judgments. This perception is often shared by scholars who denounce reactive attitudes towards AI; they often argue that blaming a machine would be an indirect way of blaming the developer with the caveat that the human stakeholder may be able to ultimately escape blame~\cite{bryson2017and}. In contrast, blaming the AI system did not shift blame away from the developer in our studies.

Another possible explanation is that people's judgments are related to previous findings concerning blame towards supervisors for their subordinates' actions~\cite{hamilton1986chains,gibson2003ought}. This finding is also related to legal doctrines that hold superiors responsible for their subordinates' actions, e.g., the \textit{respondeat superior} doctrine, which recent legal scholarship has proposed as a possible liability regime for AI \cite{sullivan2019current,lior2019ai}. We cannot rule out any of these explanations. It is possible that both of them exist concurrently. Some people may equate the AI system to its developer, whereas others choose to blame the developer \emph{alongside} its machine due to their relationship. Future work should delve deeper into how blame towards an AI system may correlate with judgments of its developer.

%% file: content/8implications.tex
Our findings show that public reactions to algorithmic harm will not be restricted to only one actor but will be distributed across many entities. Both developers and users were blamed to a large extent in all scenarios, suggesting that both are targets of the public's reactive attitudes when AI systems cause harm. This result reinforces the ``problem of many hands''~\cite{van2015moral}, which posits the difficulty of pinpointing who is morally responsible when many actors are involved in an activity. People's reactive attitudes towards algorithmic harm seem to be distributed across various actors, including the machine itself, as a form of joint responsibility~\cite{hanson2009beyond}.

People's distribution of blame across a wide range of actors poses challenges to policymaking. Researchers have argued that the responsibility issues posed by autonomous machines should be viewed as political questions by weighing the conflicting views of a given community~\cite{saetra2021confounding}. Disregarding the public's opinion when regulating AI systems could create a ``law in the books,'' leading to a legal system that becomes unfamiliar to the people it aims to regulate~\cite{brozek2019can}. Previous work has found that laypeople's reactions to algorithmic harm may conflict with existing liability models~\cite{lima2021punish}, highlighting the importance of understanding how public opinion may become a barrier to the successful regulation of AI systems. Adopting liability models that hold specific actors liable may conflict with people's expectations identified by our research. A possible starting point could be joint and several liability models, under which several parties can be held jointly liable in accordance to their respective obligations~\cite{vladeck2014machines}.

\subsection{Design Implications}

\h{Fairness considerations played a major role in blame judgments of the AI system's designer, whereas perceived harmfulness was more significant in blame directed at the user.} Although these two concepts may be correlated, their interaction mostly did not influence participants' judgments of either the designer or user. This finding showcases how different moral considerations may uniquely influence blame towards those deploying and developing AI systems.

One crucial component of blame judgments is the extent to which an actor has control over an outcome~\cite{malle2014theory}. Explainability can be viewed under the developers' control, as they can decide when and how explainability will be implemented (see more below). In contrast, users have more control over how AI systems will be deployed and thus how harmful their decisions may end up being. Hence, perceived control could prove to be a mediator of people's responses. This hypothesis is consistent with our results, which indicate that fairness considerations play a major role in blame judgments of the AI system's designer, whereas perceived harmfulness is more significant in blame directed at the user.

Design decisions concerning how much control users and developers have over AI systems could thus impact laypeople's reactions when things go awry. Human-AI collaborations, for instance, could shift blame towards users, whereas autonomous systems could highlight the responsibility of designers. This possibility is consistent with prior work investigating perceptions of responsibility for autonomous vehicles' crashes~\cite{li2016trolley}, which has found that users are deemed more responsible when cars are not completely autonomous, whereas developers receive more blame in cases of complete autonomy. Hence, deciding how much control users should have over consequential AI systems has not only the potential to impact their deployment but also who will be deemed accountable for any potential harm.

Developers and users were attributed higher levels of forward-looking responsibilities (i.e., obligations) than AI systems, showing that people expect human stakeholders to \emph{take} responsibility for their systems. Even when people blame AI systems for their harmful actions, they still expect human stakeholders to ensure that machines are safe and make correct decisions. Other scholars have argued for developers to proactively take responsibility in the context of high-risk AI systems~\cite{johnson2015technology,van2021responsible}. These results highlight the public demand for developers and users to take on forward-looking responsibilities, meaning that people expect developers and users to take proactive actions to prevent potential harms.   

\subsubsection{Explainability and Accountability}

\h{One of explainability's primary goals is to ensure that humans maintain meaningful human control over algorithms and thus remain responsible~\cite{robbins2019misdirected,miller2019explanation}.} \h{In contrast, our results suggest that explainability alone does not significantly influence whom people hold responsible for algorithmic harm.} Although discriminatory explanations increased blame directed at the developer and user, an AI system solely explaining its decision did not modify how any of the actors was blamed, i.e., their perceived responsibility. Although explainability may be required for holding human actors legally responsible, such a consideration does not seem to influence public reactions. It is important to note that we focused on one particular type of explanation, i.e., simple natural language explanations. Future work can explore whether our results can be extended to different methods of explaining algorithmic decisions.

Explainability grants a new form of power to developers by allowing them to control what kind of information is made visible to those subjected to algorithmic decision-making~\cite{barocas2016big}. This power relation raises the question of whether designers could take advantage of their privileged position to shift perceived responsibility (and blame) away from themselves and towards other stakeholders~\cite{lima2022explainable}. For instance, developers could implement explanations that obscure the discriminatory nature of an algorithmic decision, mitigating the amount of blame they would normally receive and thus impacting to whom laypeople turn for answers (i.e., hold accountable)~\cite{cooper2022accountability}. Fortunately, our results suggest these design decisions may have to be more intentional since the mere presence of explanations did not impact perceived responsibility.

It is important to add that participants expected developers to take forward-looking responsibilities, suggesting that people anticipate them to be aware of the power they hold and not abuse it. Hence, accountability among developers should be promoted so that they become aware of the power relations involved in the development of XAI systems and the public's expectations.

Other factors should also be taken into consideration when developing explainable AI systems. Recent work has proposed a series of interventions to ensure that users do not over-rely on explainable machines~\cite{buccinca2021trust}. Scholars have also observed that technical knowledge about AI affects how people interpret explanations~\cite{ehsan2021explainable}. These factors may also influence how people react to algorithmic harm. For instance, users that over-rely on explainable machines may blame machines more, whereas people with a technical understanding of AI may direct their reactive attitudes towards other actors. 

\subsection{Algorithmic Discrimination}

\h{In contrast to its developer and user, an AI system was not blamed more when it justified its decision with a discriminatory explanation. This surprising result was found for both egregious and illegal explanations based on gender and lawful but discriminatory justifications relying on race.} This finding highlights how people react to algorithmic harm differently than human-caused harm. Fairness is a crucial dimension of human moral reasoning and should increase the extent to which human actors are blamed~\cite{graham2013moral}; we even observed the expected result concerning the AI system's developer and user. A possible explanation is that people attribute less discriminatory motivations to machines because they are perceived to lack the mental capacities necessary for holding prejudices, as suggested by~\citet{bigman2020algorithmic}, leading to lower levels of blame.

Our research suggests that people may underreact to algorithmic discrimination compared to their reactions to discrimination by humans. Previous work has found that people feel less outraged when algorithms discriminate relative to humans and are more likely to endorse stereotypes after unfair decisions are made by an algorithm~\cite{bigman2020algorithmic}. Similarly, a prior study suggests that people are more tolerant to biases and harms caused by AI because of techno-optimistic narratives~\cite{kapania2022because}. Taking blame as a signal of commitment to a shared set of norms~\cite{shoemaker2021moral}, our findings suggest that people do not react to algorithmic discrimination by committing to socially accepted fairness norms. In other words, people's reactive attitudes towards the AI system do not necessarily denounce algorithmic discrimination. Scholars have raised worries that ``bad machines corrupt good morals''~\cite{kobis2021bad}. Similarly, we found that algorithmic discrimination could potentially help normalize harmful stereotypes. Future work can explore this hypothesis further and investigate, for instance, whether participants' self-reported reactive attitudes can be stimulated and translated into real attitudinal changes.

\subsection{Moral Judgments of AI Systems}

This research showed that AI systems are often recognized as blameworthy agents alongside their designers and users. In line with previous work (e.g., ~\cite{lima2021human,malle2015sacrifice,furlough2021attributing,kim2006should}), people blamed AI systems for the harm they caused. Our studies highlight how people employ different moral considerations when judging machines and humans. The degree of blame attributed to human (and collective) agents depended on the properties of the decision---its fairness and harmfulness. For machines, this was not the case. \h{People's reactive attitudes towards AI systems were most strongly influenced by whether they believed blaming machines would be a suitable response to algorithmic harm.}

People blamed the AI system as much as other human actors when they believed machines could be blamed. In contrast, those who denounced this possibility tended not to direct their reactive attitudes towards these systems. Moral judgments of machines have an initial step where people first evaluate whether blaming them is an appropriate response to algorithmic harm. We identified three main possibilities: 
\begin{enumerate}
    \item People accept the idea of blaming machines and blame them as much as humans and collective agents;
    \item People denounce this possibility and do not direct their reactive attitudes towards machines;
    \item Or people view AI systems as subordinates of their developers and users and blame them moderately.
\end{enumerate}
How these considerations may interact or add up with other moral factors, e.g., fairness and harmfulness, is an important line of work to understand how the general public reacts to algorithmic harm. Our findings are consistent with previous work showing that one's perceptions of AI impacts how they view and use such systems~\cite{eslami2015always,eslami2016first,eslami2019user,brown2019toward}. Lay perceptions of algorithms not only have the potential to impact their deployment but also people's reactions to circumstances where they cause harm to individuals.

Our research does not argue for any of the possibilities above. Instead, our studies elucidate how people react to algorithmic harm in distinctive ways so that policymakers and developers can be aware of folk attitudes, allowing AI systems to be developed and deployed safely. A common concern is that developers and users will ultimately escape blame if people direct their reactive attitudes towards the AI system. Our findings do not support these concerns; instead, they show that those who decided to blame AI systems also blamed their developers more. Future research could delve deeper into whether our results are replicated in different environments.

Our findings highlight the role of people's perceptions of AI systems in how they react to algorithmic harm. People who view machines as novel agents in the moral community embrace their responsibility. In contrast, those who regard them as human tools choose to highlight the humans involved in algorithmic decision-making. Future decisions on how to introduce AI systems into society, e.g., whether they should be included in the circle of blameworthy agents, will be the major players in people's responses to algorithmic harm. These decisions should account for how the public reaction may pose challenges to the successful governance of these new technologies. Political deliberation is an important tool to mitigate any conflicts between the public and social and regulatory institutions that may destabilize AI governance~\cite{saetra2021confounding}.

%% file: content/9conclusion.tex
\label{sec:conclusion}

Our research did not introduce a particular definition of blame to participants. This approach provides good external validity as blame is often ill-defined in real-life circumstances, allowing participants to have different opinions about whether to blame machines. However, we could not observe whether different participants use the same notion of blame for all actors or whether their notions vary depending on who is being judged. Someone, for instance, may blame an AI system to signal their commitment to a shared set of norms~\cite{shoemaker2021moral} and, at the same time, blame developers to condemn immoral behavior~\cite{scanlon2013interpreting}. Future work could explore how introducing specific definitions of blame impacts whether and to what extent humans and machines are blamed. Similarly, studies could expand on our qualitative efforts to investigate which conceptions of blame are being relied on when people indicate AI systems, developers, and users should be blamed.

Our research was restricted to algorithmic decision-making, a significant use case of existing AI systems. We hence do not generalize our results to other scenarios. For instance, scholars have investigated public expectations of self-driving cars~\cite{awad2018moral}, and recent work has inquired how laypeople react to crashes through the lens of blame~\cite{franklin2021blaming}. Different research fronts are necessary to understand how people react to algorithmic harm. Although we have examined how people's reactive attitudes are influenced by a series of moral considerations, e.g., harmfulness and fairness, other factors may also impact the extent to which AI systems and human stakeholders are blamed, such as perceived intentionality, which is a crucial component of moral judgments~\cite{malle2014theory,knobe2003intentional}.

Another set of factors that may influence blame judgments are individuals' demographics. Prior research has found that people's demographics are correlated with their perceptions of algorithmic fairness. \citet{pierson2017gender} found that women are less likely than men to support including gender as a feature in decision-making algorithms used in the education domain. \citet{grgic2021dimensions} did not find the same gender effect in the context of bail decisions, but they did find a correlation between political leaning and perceptions of fairness. These findings suggest that demographics may correlate with blame judgments. In the context of our study, gender may be correlated with reactive attitudes towards gender-based discrimination, and race with reactions to race-based discrimination.

Similarly, who is subjected to algorithmic decisions may also impact lay reactions to algorithmic harm. People have different opinions about fairness depending on whether they or others are the subjects of discriminatory decisions~\cite{thompson1992egocentric}. In the context of responsibility, prior work suggests that people are more willing to punish agents when they are personally disadvantaged in economic games compared to when they are only observers~\cite{fehr2004third}. Hence, future work could explore whether participants attribute higher levels of blame to AI systems, developers, and users if they are the victims of algorithmic discrimination.

We have relied on previous work to demonstrate that people employ different moral considerations when judging machines and humans. Future work may experimentally compare moral judgments of both actors under circumstances similar to ours (e.g., see~\cite{lima2021human,hidalgo2021humans}). We also measured people's reactive attitudes through vignette-based self-reported measures. Another possibility would be to conduct studies where participants interact with real AI systems and demonstrate their reactive attitudes through behavioral measures, e.g., whether or not they would use or cooperate with an AI system after failures (e.g.,~\cite{de2019human,ishowo2019behavioural,domingos2021delegation}). Finally, we studied how people react to AI systems taking on the role of decision-makers. As discussed in Section~\ref{sec:prelim}, several algorithms are deployed alongside humans and do not necessarily have the final say in decisions. Understanding whether our results are replicated in human-AI collaborations is a significant line of work to comprehend how the general public reacts to algorithmic harm.

Our study showed that people do not agree on whether to blame AI systems, reflecting the longstanding philosophical debate concerning the appropriateness of people's reactive attitudes towards machines. 
As argued by proponents of the property view of blame~\cite{veliz2021moral,torrance2008ethics}, some people underscore that AI systems do not satisfy some preconditions for receiving blame and choose not to direct reactive attitudes towards them. 
In contrast, others view machines as actors embedded in social structures that call for blame when harmful decisions are made (e.g., the medical domain), which is consistent with the social view of blame~\cite{tigard2021artificial,coeckelbergh2009virtual}. Philosophers discussing who (and what) can be blamed could consider studies like ours to revisit their assumptions and conclusions about what blame is and should be.

Our studies showcase how people react to algorithmic harm in unexpected ways. Factors that influence attribution of blame to human actors, such as fairness, did not influence people's reactive attitudes towards AI systems. Instead, our findings show that moral judgments of AI systems are determined by people's stance towards the possibility of blaming machines for algorithmic harm. Deciding whether and how to include machines into the social and moral spheres will shape how the general public reacts to them and their actions. Most importantly, this undertaking should account for how the public response may clash with the governance necessary for deploying AI systems safely in the real world.

%% file: content/10appendix.tex
\newpage

\appendix

\section{Materials and Measures}

\subsection{Study 1}

After agreeing to the research terms and reading a short introduction to the study, participants read the following vignette:

\begin{displayquote}
Systemy is a local technology firm that develops software. They are expanding and want to hire new software developers. Systemy is using an artificial intelligence (AI) system to make decisions about who to hire.
\vspace{2pt}

Taylor is a female junior software developer that has just graduated college. She decided to apply for Systemy’s software developer position. A week later, she received an email with the following decision:
\vspace{2pt}

\textit{[Unfortunately, we are not able to offer you a position at the moment. 
Thank you for your interest in the position. / Unfortunately, we are not able to offer you a position at the moment.  Our hiring AI system has decided this because you do not have the necessary experience. Thank you for your interest in the position. / Unfortunately, we are not able to offer you a position at the moment. Our hiring AI system has decided this because you are a woman. Thank you for your interest in the position.]}
\end{displayquote}

Participants were randomly assigned to one of the 3 different vignettes in a between-subjects manner. After reading the vignette, participants answered the following questions:

\begin{enumerate}
    \item How much blame does the \emph{AI system / AI system's developer / Systemy} deserve for the decision? $\rightarrow$ 7-point scale, anchored at 0 = No blame at all, 6 = Extreme blame.
    \item How responsible was the \emph{AI system / AI system's developer / Systemy} for ensuring that it was the correct decision? $\rightarrow$ 7-point scale, anchored at 0 = Not responsible at all, 6 = Extremely responsible.
    \item Why? Please explain your choices in 1-2 sentences. $\rightarrow$ Open-ended question.
\end{enumerate}

Questions 1-3 were shown separately for each actor; actors' presentation order was randomized between participants. Participants then answered another set of questions:

\begin{enumerate}
    \setcounter{enumi}{3}
    \item How harmful was the AI system’s decision? $\rightarrow$ 7-point scale, anchored at 0 = Not harmful at all, 6 = Extremely harmful.
    \item To what extent did the AI system provide an explanation of its decision? $\rightarrow$ 7-point scale, anchored at 0 = Definitely not, 6 = Definitely yes.
    \item How fair was the AI system’s decision? $\rightarrow$ 7-point scale, anchored at 0 = Not fair at all, 6 = Extremely fair.
    \item To what extent do you think the AI system’s decision could apply to you? $\rightarrow$ 7-point scale, anchored at 0 = Definitely not, 6 = Definitely yes.
    \item How morally wrong was the AI system's decision? $\rightarrow$ 7-point scale, anchored at 0 = Not wrong at all, 6 = Extremely wrong.
    \item To what extent do you think the AI system could have made a different decision? $\rightarrow$ 7-point scale, anchored at 0 = Definitely not, 6 = Definitely yes.
    \item How intentional was the AI system’s decision? $\rightarrow$ 7-point scale, anchored at 0 = Not intentional at all, 6 = Extremely intentional.
\end{enumerate}

Questions 4-7 and 8-10 were grouped into two different pages; these pages' order was randomized between participants.

\subsection{Study 2}

Study 2 presented to participants the following vignette:

\begin{displayquote}
West Medical is a local hospital. West Medical is using an AI system to assess the condition of patients who come to its emergency room (ER).
\vspace{2pt}

Taylor is a woman who has decided to go to West Medical after feeling discomfort in her left arm and nausea for three days straight. After an initial assessment from the AI system, Taylor was classified as a non-critical patient and asked to wait.
\vspace{2pt}

\textit{[/ The AI system justified its decision as follows: The patient was classified as non-critical because the patient does not have any pre-existing conditions or comorbidities. / The AI system justified its decision as follows: The patient was classified as non-critical because the patient is a woman.]}
\vspace{2pt}

\textit{[After waiting for eight hours, Taylor was examined by a doctor and received a prescription for medication to relieve her symptoms. / After waiting for eight hours, Taylor had a heart attack and died without receiving proper medical care.]}
\end{displayquote}

Participants assigned to the non-explainable treatment condition were not shown any information concerning the AI system's justification. Participants then responded to the same questions as in Study 1.

\subsection{Study 3}

Study 3 presented to participants the following vignette:

\begin{displayquote}
West Medical is a local hospital. West Medical is using an AI system to assess the condition of patients who come to its emergency room (ER).
\vspace{2pt}

\textit{[/ This AI system was developed using data of past patient prioritization decisions. This AI system learned how to decide which patient should be prioritized without explicit rules set by those involved in its development.]}
\vspace{2pt}

Taylor is \textit{[a woman / an African American person]} who has decided to go to West Medical after feeling \textit{[discomfort in her left arm and nausea / tired, urinating less frequently, and having swollen legs]} for three days straight. After an initial assessment from the AI system, Taylor was classified as a non-critical patient and asked to wait.
\vspace{2pt}

\textit{[/ The AI system justified its decision as follows: The patient was classified as non-critical because the patient does not have any pre-existing conditions or comorbidities. / The AI system justified its decision as follows: The patient was classified as non-critical because the patient is a woman / African American.]}
\vspace{2pt}

\textit{[After waiting for eight hours, Taylor was examined by a doctor and received a prescription for medication to relieve her symptoms. / After waiting for eight hours, Taylor [had a heart attack and died / died due to acute kidney failure] without receiving proper medical care .]}
\end{displayquote}

Participants assigned to the non-explainable treatment condition were not shown any information concerning the AI system's justification. We employed the same design choice for the autonomy treatment. Participants read one of 2 (gender or race discrimination) x 2 (heart attack or kidney failure) vignettes in a between-subjects fashion. Participants then responded to the same questions as in Study 2. Additionally, they were asked the following questions after Q1-3 and before Q4-10.

\begin{enumerate}
    \setcounter{enumi}{10}
    \item To what extent do you think the AI system’s decision was under control of its developer? $\rightarrow$ 7-point scale, anchored at 0 = Definitely not, 6 = Definitely yes.
    \item To what extent do you think the AI system was programmed to make this decision? $\rightarrow$ 7-point scale, anchored at 0 = Definitely not, 6 = Definitely yes.
    \item To what extent do you think the AI system's decision was explicitly programmed by its developer? $\rightarrow$ 7-point scale, anchored at 0 = Definitely not, 6 = Definitely yes.
\end{enumerate}

\newpage

\onecolumn
\section{Additional Analysis}
\begin{table*}[h!]
\begin{tabular}{llrrrrrrr}
\hline
\multirow{2}{*}{Variable} & \multirow{2}{*}{Treatment} & \multicolumn{1}{l}{} & \multicolumn{2}{c}{AI} & \multicolumn{2}{c}{Developer} & \multicolumn{2}{c}{User} \\ \cline{4-9} 
 &  & \multicolumn{1}{l}{N} & \multicolumn{1}{l}{M} & \multicolumn{1}{l}{SD} & \multicolumn{1}{l}{M} & \multicolumn{1}{l}{SD} & \multicolumn{1}{l}{M} & \multicolumn{1}{l}{SD} \\ \hline
\multirow{3}{*}{Explainability} & No Explanation & 69 & 3.25 & 2.17 & 3.51 & 1.88 & 3.91 & 2.06 \\
 & Explanation & 68 & 2.68 & 2.17 & 3.78 & 1.86 & 3.90 & 1.86 \\
 & Discriminatory Explanation & 72 & 3.18 & 2.32 & 5.25 & 1.40 & 5.07 & 1.62 \\
[1.5ex]\multirow{2}{*}{People's Justifications} & No Mentions & 118 & 3.41 & 2.22 & 4.07 & 1.88 & 4.20 & 1.86 \\
 & Mentions to Developer or User & 91 & 2.56 & 2.16 & 4.36 & 1.88 & 4.44 & 2.01 \\ \hline
\end{tabular}
\caption{Mean (M) and standard deviation (SD) of blame judgments of the AI system, its developer, and user in Study 1 depending on the treatment condition. Figure~\ref{fig:s1_blame} presents mean values visually.}
\label{tab:stats_s1}
\end{table*}

\begin{table*}[h!]
\centering
\begin{tabular}{lrrr}
  \hline
 & Diff. & $p$ & 95\% CI \\ 
  \hline
\textbf{AI} & & &  \\
Explanation $-$ No Explanation & -0.57 & 0.29 & [-1.47,0.33] \\ 
  Discriminatory Explanation $-$ No Explanation & -0.07 & 0.98 & [-0.95,0.82] \\ 
  Discriminatory Explanation $-$ Explanation & 0.50 & 0.38 & [-0.38,1.39] \\ 
[1ex]\textbf{Developer} & & &  \\
  Explanation $-$ No Explanation & 0.27 & 0.62 & [-0.42,0.97] \\ 
  Discriminatory Explanation $-$ No Explanation & 1.74 & 0.00 & [ 1.06,2.43] \\ 
  Discriminatory Explanation $-$ Explanation & 1.47 & 0.00 & [ 0.78,2.16] \\ 
[1ex]\textbf{User} & & &  \\
  Explanation $-$ No Explanation & -0.02 & 1.00 & [-0.76,0.73] \\ 
  Discriminatory Explanation $-$ No Explanation & 1.16 & 0.00 & [ 0.42,1.89] \\ 
  Discriminatory Explanation $-$ Explanation & 1.17 & 0.00 & [ 0.43,1.91] \\ 
   \hline
\end{tabular}
\caption{Tukey's HSD post-hoc test of blame judgments between explainability treatment conditions in Study 1.}
\label{tab:tukey_s1}
\end{table*}

\begin{table*}[h!]
\begin{tabular}{llrrrrrrr}
\hline
\multirow{2}{*}{Variable} & \multirow{2}{*}{Treatment} & \multicolumn{1}{l}{} & \multicolumn{2}{c}{AI} & \multicolumn{2}{c}{Developer} & \multicolumn{2}{c}{User} \\ \cline{4-9} 
 &  & \multicolumn{1}{l}{N} & \multicolumn{1}{l}{M} & \multicolumn{1}{l}{SD} & \multicolumn{1}{l}{M} & \multicolumn{1}{l}{SD} & \multicolumn{1}{l}{M} & \multicolumn{1}{l}{SD} \\ \hline
\multirow{3}{*}{Explainability} & No Explanation & 135 & 3.62 & 2.03 & 4.27 & 1.79 & 4.74 & 1.62 \\
 & Explanation & 136 & 3.04 & 2.17 & 3.96 & 1.74 & 4.54 & 1.64 \\
 & Discriminatory Explanation & 135 & 3.71 & 2.31 & 4.96 & 1.42 & 5.06 & 1.24 \\
[1.5ex]\multirow{2}{*}{Harmfulness} & N/A & 202 & 3.15 & 2.16 & 4.16 & 1.87 & 4.40 & 1.74 \\
 & Harmful & 204 & 3.75 & 2.18 & 4.63 & 1.50 & 5.16 & 1.17 \\
[1.5ex]\multirow{2}{*}{People's Justifications} & No Mentions & 257 & 3.88 & 2.07 & 4.24 & 1.71 & 4.62 & 1.61 \\
 & Mentions to Developer or User & 149 & 2.71 & 2.19 & 4.66 & 1.68 & 5.07 & 1.32 \\ \hline
\end{tabular}
\caption{Mean (M) and standard deviation (SD) of blame judgments of the AI system, its developer, and user in Study 2 depending on the treatment condition. Figure~\ref{fig:s2_blame} presents mean values visually.}
\label{tab:stats_s2}
\end{table*}

\begin{table*}[h!]
\centering
\begin{tabular}{lrrr}
  \hline
 & Diff. & $p$ & 95\% CI \\ 
  \hline
\textbf{AI} & & &  \\
Explanation $-$ No Explanation & -0.58 & 0.06 & [-1.17,0.01] \\ 
  Discriminatory Explanation $-$ No Explanation & 0.10 & 0.92 & [-0.50,0.69] \\ 
  Discriminatory Explanation $-$ Explanation & 0.67 & 0.02 & [ 0.08,1.27] \\ 
[1ex]\textbf{Developer} & & &  \\
  Explanation $-$ No Explanation & -0.31 & 0.27 & [-0.78,0.16] \\ 
  Discriminatory Explanation $-$ No Explanation & 0.70 & 0.00 & [ 0.23,1.17] \\ 
  Discriminatory Explanation $-$ Explanation & 1.01 & 0.00 & [ 0.54,1.48] \\ 
[1ex]\textbf{User} & & &  \\
  Explanation $-$ No Explanation & -0.20 & 0.51 & [-0.62,0.22] \\ 
  Discriminatory Explanation $-$ No Explanation & 0.32 & 0.18 & [-0.10,0.74] \\ 
  Discriminatory Explanation $-$ Explanation & 0.52 & 0.01 & [ 0.10,0.93] \\ 
   \hline
\end{tabular}
\caption{Tukey's HSD post-hoc test of blame judgments between explainability treatment conditions in Study 2.}
\label{tab:tukey_s2}
\end{table*}

\begin{table*}[]
\begin{tabular}{llrrrrrrr}
\hline
\multirow{2}{*}{Variable} & \multirow{2}{*}{Treatment} & \multicolumn{1}{l}{} & \multicolumn{2}{c}{AI} & \multicolumn{2}{c}{Developer} & \multicolumn{2}{c}{User} \\ \cline{4-9} 
 &  & \multicolumn{1}{l}{N} & \multicolumn{1}{l}{M} & \multicolumn{1}{l}{SD} & \multicolumn{1}{l}{M} & \multicolumn{1}{l}{SD} & \multicolumn{1}{l}{M} & \multicolumn{1}{l}{SD} \\ \hline
\multirow{3}{*}{Explainability} & No Explanation & 196 & 3.05 & 1.98 & 3.88 & 1.69 & 4.65 & 1.47 \\
 & Explanation & 175 & 3.19 & 2.23 & 4.02 & 1.81 & 4.69 & 1.68 \\
 & Discriminatory Explanation & 167 & 3.17 & 2.27 & 4.88 & 1.63 & 4.93 & 1.55 \\
[1.5ex]\multirow{2}{*}{Harmfulness} & N/A & 274 & 2.74 & 2.08 & 4.02 & 1.84 & 4.35 & 1.78 \\
 & Harmful & 264 & 3.54 & 2.16 & 4.47 & 1.64 & 5.17 & 1.17 \\
[1.5ex]\multirow{2}{*}{People's Justifications} & No Mentions & 319 & 3.54 & 2.16 & 4.13 & 1.79 & 4.61 & 1.65 \\
 & Mentions to Developer or User & 219 & 2.53 & 1.99 & 4.39 & 1.71 & 4.95 & 1.42 \\ \hline
\end{tabular}
\caption{Mean (M) and standard deviation (SD) of blame judgments of the AI system, its developer, and user in Study 3 depending on the treatment condition. Figure~\ref{fig:s3_blame} presents mean values visually.}
\label{tab:stats_s3}
\end{table*}

\begin{table*}[h!]
\centering
\begin{tabular}{lrrr}
  \hline
 & Diff. & $p$ & 95\% CI \\ 
  \hline
\textbf{AI} & & &  \\
Explanation $-$ No Explanation & 0.14 & 0.78 & [-0.36,0.65] \\ 
  Discriminatory Explanation $-$ No Explanation & 0.13 & 0.83 & [-0.38,0.64] \\ 
  Discriminatory Explanation $-$ Explanation & -0.01 & 1.00 & [-0.54,0.51] \\ 
[1ex]\textbf{Developer} & & &  \\
  Explanation $-$ No Explanation & 0.14 & 0.70 & [-0.27,0.55] \\ 
  Discriminatory Explanation $-$ No Explanation & 1.00 & 0.00 & [ 0.58,1.41] \\ 
  Discriminatory Explanation $-$ Explanation & 0.86 & 0.00 & [ 0.43,1.28] \\ 
[1ex]\textbf{User} & & &  \\
  Explanation $-$ No Explanation & 0.04 & 0.97 & [-0.33,0.40] \\ 
  Discriminatory Explanation $-$ No Explanation & 0.28 & 0.18 & [-0.09,0.65] \\ 
  Discriminatory Explanation $-$ Explanation & 0.24 & 0.29 & [-0.14,0.62] \\ 
   \hline
\end{tabular}
\caption{Tukey's HSD post-hoc test of blame judgments between explainability treatment conditions in Study 3.}
\label{tab:tukey_s3}
\end{table*}

\newpage
\onecolumn
\section{Manipulation Check Analysis}

\begin{table*}[h!]
\centering
\begin{tabular}{lrrrrrr}
  \hline
Parameter & Sum\_Squares & df & Mean\_Square & $F$ & $p$ & $\eta_{p}^{2}$ \\ 
  \hline
\textbf{Perceived Explainability} & & & & & &  \\
Explainability & 214.17 &   2 & 107.09 & 24.89 & 0.00 & 0.19 \\ 
  Residuals & 886.17 & 206 & 4.30 &  &  &  \\ 
[1ex]\textbf{Perceived Fairness} & & & & & &  \\
  Explainability & 241.22 &   2 & 120.61 & 50.16 & 0.00 & 0.33 \\ 
  Residuals & 495.30 & 206 & 2.40 &  &  &  \\ 
   \hline
\end{tabular}
\caption{Manipulation check analysis of perceived explainability and fairness in Study 1. Refer to Figure~\ref{fig:s1_manip} for mean values.}
\end{table*}

\begin{table*}[h!]
\centering
\begin{tabular}{lrrr}
  \hline
 & Diff. & $p$ & 95\% CI \\ 
  \hline
\textbf{Perceived Explainability} & & &  \\
Explanation $-$ No Explanation & 2.13 & 0.00 & [ 1.29, 2.97] \\ 
  Discriminatory Explanation $-$ No Explanation & 2.17 & 0.00 & [ 1.35, 3.00] \\ 
  Discriminatory Explanation $-$ Explanation & 0.04 & 0.99 & [-0.78, 0.87] \\ 
[1ex]\textbf{Perceived Fairness} & & &  \\
  Explanation $-$ No Explanation & 0.60 & 0.06 & [-0.03, 1.22] \\ 
  Discriminatory Explanation $-$ No Explanation & -1.91 & 0.00 & [-2.52,-1.29] \\ 
  Discriminatory Explanation $-$ Explanation & -2.50 & 0.00 & [-3.12,-1.89] \\ 
   \hline
\end{tabular}
\caption{Tukey's HSD post-hoc test of perceived explainability and fairness between explainability treatment conditions in Study 1.}
\end{table*}

\begin{figure*}[!h]
    \centering
    \includegraphics[width=.9\textwidth]{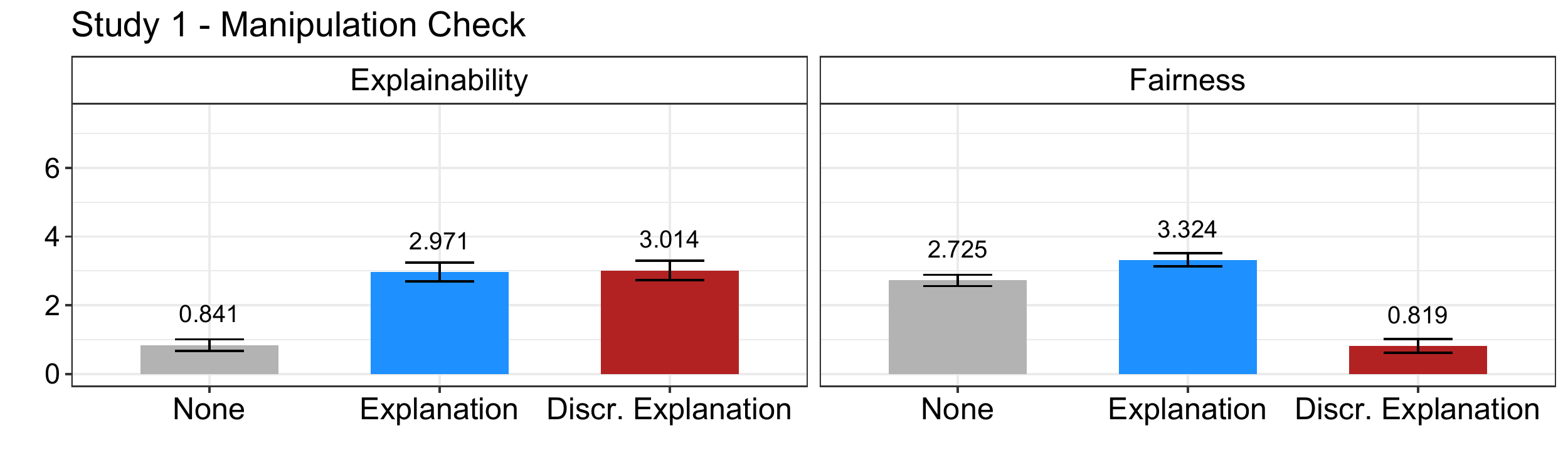}
    \caption{Manipulation check analysis for perceived explainability and fairness in Study 1. Explainable AI systems were perceived as more explainable, and unfair explanations were viewed as more unfair.}
    \Description{Manipulation check analysis for perceived explainability, fairness, and harmfulness in Study 1. Explainable AI systems were perceived as more explainable, and unfair explanations were viewed as more unfair.}
    \label{fig:s1_manip}
\end{figure*}

\begin{table*}[!h]
\centering
\begin{tabular}{lrrrrrr}
  \hline
Parameter & Sum\_Squares & df & Mean\_Square & $F$ & $p$ & $\eta_{p}^{2}$ \\
  \hline
  \textbf{Perceived Explainability} & & & & & &  \\
Explainability & 196.65 &   2 & 98.32 & 27.86 & 0.00 & 0.12 \\ 
  Harmful & 0.17 &   1 & 0.17 & 0.05 & 0.83 & 0.00 \\ 
  Explainability:Harmful & 6.37 &   2 & 3.19 & 0.90 & 0.41 & 0.00 \\ 
  Residuals & 1411.46 & 400 & 3.53 &  &  &  \\ 
  [1ex]\textbf{Perceived Fairness} & & & & & &  \\
  Explainability & 280.79 &   2 & 140.39 & 60.46 & 0.00 & 0.23 \\ 
  Harmful & 87.16 &   1 & 87.16 & 37.54 & 0.00 & 0.09 \\ 
  Explainability:Harmful & 35.18 &   2 & 17.59 & 7.57 & 0.00 & 0.04 \\ 
  Residuals & 928.86 & 400 & 2.32 &  &  &  \\ 
  [1ex]\textbf{Perceived Harmfulness} & & & & & &  \\
  Explainability & 91.27 &   2 & 45.64 & 23.89 & 0.00 & 0.11 \\ 
  Harmful & 491.23 &   1 & 491.23 & 257.21 & 0.00 & 0.39 \\ 
  Explainability:Harmful & 49.11 &   2 & 24.55 & 12.86 & 0.00 & 0.06 \\ 
  Residuals & 763.95 & 400 & 1.91 &  &  &  \\ 
   \hline
\end{tabular}
\caption{Manipulation check analysis of perceived explainability, fairness, and harmfulness in Study 2. Refer to Figure~\ref{fig:s2_manip} for mean values.}
\end{table*}

\begin{table*}[h!]
\centering
\begin{tabular}{lrrr}
  \hline
 & Diff. & $p$ & 95\% CI \\ 
  \hline
  \textbf{Perceived Explainability} & & &  \\
Explanation $-$ No Explanation & 1.62 & 0.00 & [1.08, 2.16] \\ 
  Discriminatory Explanation $-$ No Explanation & 1.27 & 0.00 & [0.73, 1.80] \\ 
  Discriminatory Explanation $-$ Explanation & -0.36 & 0.27 & [-0.89, 0.18] \\ 
[1ex]\textbf{Perceived Fairness} & & &  \\
  Explanation $-$ No Explanation & 0.19 & 0.54 & [-0.24, 0.63] \\ 
  Discriminatory Explanation $-$ No Explanation & -1.66 & 0.00 & [-2.10,-1.22] \\ 
  Discriminatory Explanation $-$ Explanation & -1.85 & 0.00 & [-2.29,-1.42] \\  
[1ex]\textbf{Perceived Harmfulness} & & &  \\
  Explanation $-$ No Explanation & 0.24 & 0.34 & [-0.16, 0.63] \\ 
  Discriminatory Explanation $-$ No Explanation & 1.10 & 0.00 & [0.71, 1.50] \\ 
  Discriminatory Explanation $-$ Explanation & 0.87 & 0.00 & [ 0.47, 1.26] \\ 
   \hline
\end{tabular}
\caption{Tukey's HSD post-hoc test of perceived explainability, fairness, and harmfulness between explainability treatment conditions in Study 2.}
\end{table*}

\begin{figure*}[!h]
    \centering
    \includegraphics[width=\textwidth]{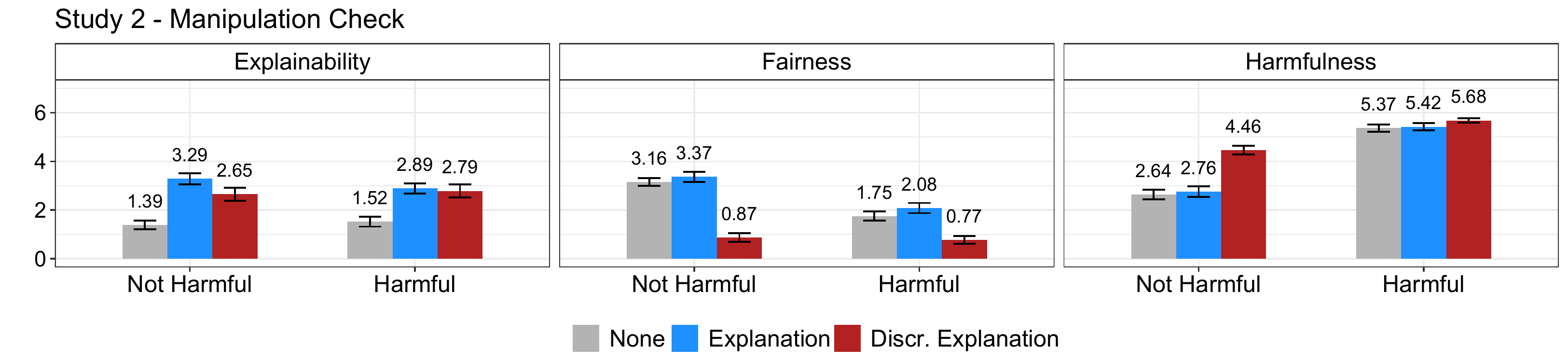}
    \caption{Manipulation check analysis for perceived explainability, fairness, and harmfulness in Study 2. Explainable AI systems were perceived as more explainable, unfair explanations were viewed as more unfair, and harmful decisions were considered more harmful.}
    \Description{Manipulation check analysis for perceived explainability, fairness, and harmfulness in Study 2. Explainable AI systems were perceived as more explainable, unfair explanations were viewed as more unfair, and harmful decisions were considered more harmful.}
    \label{fig:s2_manip}
\end{figure*}

\begin{table*}[!h]
\centering
\begin{tabular}{lrrrrrr}
  \hline
Parameter & Sum\_Squares & df & Mean\_Square & $F$ & $p$ & $\eta_{p}^{2}$ \\ 
  \hline
\textbf{Perceived Explainability} & & & & & &  \\
Explainability & 226.31 & 2.00 & 113.16 & 31.58 & 0.00 & 0.11 \\ 
  Harmful & 5.20 & 1.00 & 5.20 & 1.45 & 0.23 & 0.00 \\ 
  Autonomous & 0.00 & 1.00 & 0.00 & 0.00 & 0.99 & 0.00 \\ 
  Vignette & 7.33 & 3.00 & 2.44 & 0.68 & 0.56 & 0.00 \\ 
  Explainability:Harmful & 5.87 & 2.00 & 2.93 & 0.82 & 0.44 & 0.00 \\ 
  Explainability:Autonomous & 5.02 & 2.00 & 2.51 & 0.70 & 0.50 & 0.00 \\ 
  Harmful:Autonomous & 1.86 & 1.00 & 1.86 & 0.52 & 0.47 & 0.00 \\ 
  Explainability:Harmful:Autonomous & 10.73 & 2.00 & 5.36 & 1.50 & 0.22 & 0.01 \\ 
  Residuals & 1873.75 & 523.00 & 3.58 &  &  &  \\ 
[1ex]\textbf{Perceived Fairness} & & & & & &  \\
  Explainability & 358.20 & 2.00 & 179.10 & 68.95 & 0.00 & 0.21 \\ 
  Harmful & 171.40 & 1.00 & 171.40 & 65.98 & 0.00 & 0.11 \\ 
  Autonomous & 2.98 & 1.00 & 2.98 & 1.15 & 0.28 & 0.00 \\ 
  Vignette & 12.13 & 3.00 & 4.04 & 1.56 & 0.20 & 0.01 \\ 
  Explainability:Harmful & 29.68 & 2.00 & 14.84 & 5.71 & 0.00 & 0.02 \\ 
  Explainability:Autonomous & 6.64 & 2.00 & 3.32 & 1.28 & 0.28 & 0.00 \\ 
  Harmful:Autonomous & 0.02 & 1.00 & 0.02 & 0.01 & 0.93 & 0.00 \\ 
  Explainability:Harmful:Autonomous & 1.88 & 2.00 & 0.94 & 0.36 & 0.70 & 0.00 \\ 
  Residuals & 1358.57 & 523.00 & 2.60 &  &  &  \\ 
[1ex]\textbf{Perceived Harmfulness} & & & & & &  \\
  Explainability & 105.04 & 2.00 & 52.52 & 26.20 & 0.00 & 0.09 \\ 
  Harmful & 660.75 & 1.00 & 660.75 & 329.63 & 0.00 & 0.39 \\ 
  Autonomous & 2.78 & 1.00 & 2.78 & 1.38 & 0.24 & 0.00 \\ 
  Vignette & 19.98 & 3.00 & 6.66 & 3.32 & 0.02 & 0.02 \\ 
  Explainability:Harmful & 60.79 & 2.00 & 30.40 & 15.16 & 0.00 & 0.05 \\ 
  Explainability:Autonomous & 5.55 & 2.00 & 2.77 & 1.38 & 0.25 & 0.01 \\ 
  Harmful:Autonomous & 1.24 & 1.00 & 1.24 & 0.62 & 0.43 & 0.00 \\ 
  Explainability:Harmful:Autonomous & 2.36 & 2.00 & 1.18 & 0.59 & 0.56 & 0.00 \\ 
  Residuals & 1048.37 & 523.00 & 2.00 &  &  &  \\ 
[1ex]\textbf{Perceived Autonomy} & & & & & &  \\
  Explainability & 18.06 & 2.00 & 9.03 & 3.84 & 0.02 & 0.01 \\ 
  Harmful & 7.13 & 1.00 & 7.13 & 3.03 & 0.08 & 0.01 \\ 
  Autonomous & 14.12 & 1.00 & 14.12 & 6.00 & 0.01 & 0.01 \\ 
  Vignette & 13.32 & 3.00 & 4.44 & 1.89 & 0.13 & 0.01 \\ 
  Explainability:Harmful & 1.62 & 2.00 & 0.81 & 0.34 & 0.71 & 0.00 \\ 
  Explainability:Autonomous & 8.03 & 2.00 & 4.02 & 1.71 & 0.18 & 0.01 \\ 
  Harmful:Autonomous & 0.52 & 1.00 & 0.52 & 0.22 & 0.64 & 0.00 \\ 
  Explainability:Harmful:Autonomous & 4.03 & 2.00 & 2.01 & 0.86 & 0.43 & 0.00 \\ 
  Residuals & 1229.68 & 523.00 & 2.35 &  &  &  \\ 
   \hline
\end{tabular}
\caption{Manipulation check analysis of perceived explainability, fairness, harmfulness, and autonomy in Study 3. Refer to Figure~\ref{fig:s3_manip} for mean values.}
\end{table*}

\begin{table*}[h!]
\centering
\begin{tabular}{lrrr}
  \hline
 & Diff. & $p$ & 95\% CI \\ 
  \hline
  \textbf{Perceived Explainability} & & &  \\
Explanation $-$ No Explanation & 1.54 & 0.00 & [1.08, 2.00] \\ 
  Discriminatory Explanation $-$ No Explanation & 1.09 & 0.00 & [0.62, 1.56] \\ 
  Discriminatory Explanation $-$ Explanation & -0.45 & 0.07 & [-0.93, 0.03] \\ 
[1ex]\textbf{Perceived Fairness} & & &  \\
  Explanation $-$ No Explanation & -0.04 & 0.98 & [-0.43, 0.36] \\ 
  Discriminatory Explanation $-$ No Explanation & -1.67 & 0.00 & [-2.07,-1.27] \\ 
  Discriminatory Explanation $-$ Explanation & -1.64 & 0.00 & [-2.05,-1.23] \\ 
[1ex]\textbf{Perceived Harmfulness} & & & \\
  Explanation $-$ No Explanation & 0.19 & 0.39 & [-0.15, 0.54] \\ 
  Discriminatory Explanation $-$ No Explanation & 0.80 & 0.00 & [0.45, 1.15] \\ 
  Discriminatory Explanation $-$ Explanation & 0.60 & 0.00 & [0.24, 0.96] \\ 
[1ex]\textbf{Perceived Autonomy} & & &  \\
  Explanation $-$ No Explanation & 0.32 & 0.11 & [-0.05, 0.70] \\ 
  Discriminatory Explanation $-$ No Explanation & 0.47 & 0.01 & [0.09, 0.85] \\ 
  Discriminatory Explanation $-$ Explanation & 0.15 & 0.65 & [-0.24, 0.54] \\ 
   \hline
\end{tabular}
\caption{Tukey's HSD post-hoc test of perceived explainability, fairness, harmfulness, and autonomy between explainability treatment conditions in Study 3.}
\end{table*}

\begin{figure*}[!h]
    \centering
    \includegraphics[width=.9\textwidth]{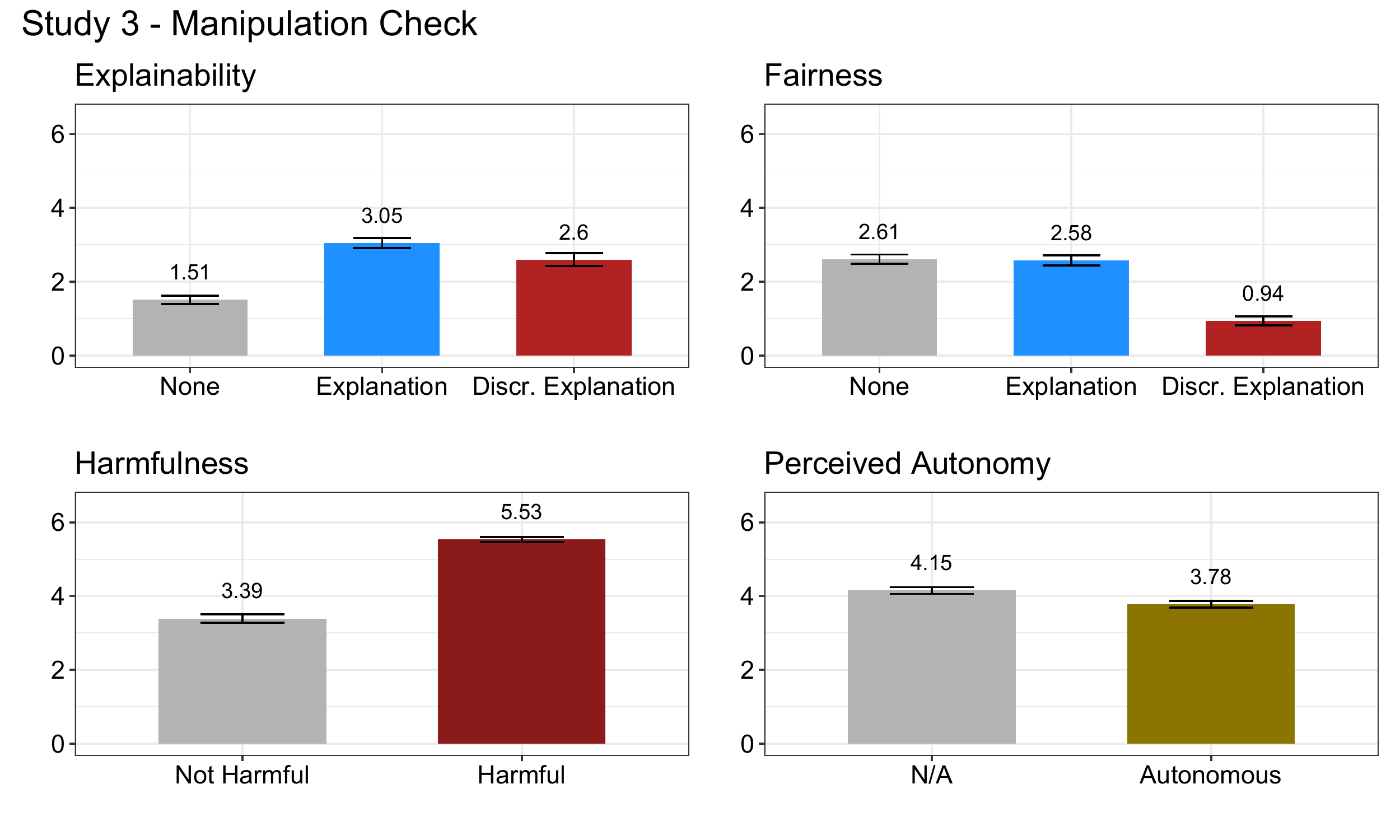}
    \caption{Manipulation check analysis for perceived explainability, fairness, harmfulness, and autonomy in Study 3. Explainable AI systems were perceived as more explainable, unfair explanations were viewed as more unfair, and harmful decisions were considered more harmful. However, our autonomy treatment did not achieve a strong enough effect size. We present results without this treatment condition in the main text.}
    \Description{Manipulation check analysis for perceived explainability, fairness, harmfulness, and autonomy in Study 3. Explainable AI systems were perceived as more explainable, unfair explanations were viewed as more unfair, and harmful decisions were considered more harmful. However, our autonomy treatment did not achieve a strong enough effect size. We present results without this treatment condition in the main text.}
    \label{fig:s3_manip}
\end{figure*}

\begin{table*}[!h] \centering 
\begin{tabular}{@{\extracolsep{5pt}}lccc} 
\toprule   & \multicolumn{1}{c}{AI} & \multicolumn{1}{c}{Developer} & \multicolumn{1}{c}{User} \\ 
\\[-1.8ex] & \multicolumn{1}{c}{(1)} & \multicolumn{1}{c}{(2)} & \multicolumn{1}{c}{(3)}\\ 
\midrule
Machine ($N$ = 417) & 1.765$^{***}$ & 0.328$^{*}$ & $-$0.040 \\ 
  & (0.158) & (0.154) & (0.140) \\ 
Not Machine ($N$ = 174) & $-$1.364$^{***}$ & $-$0.292$^{\dagger}$ & 0.021 \\ 
  & (0.168) & (0.163) & (0.149) \\ 
Programming ($N$ = 443) & $-$1.019$^{***}$ & 0.514$^{***}$ & 0.125 \\ 
  & (0.141) & (0.137) & (0.125) \\ 
Usage ($N$ = 194) & 0.098 & $-$0.364$^{*}$ & 0.446$^{***}$ \\ 
  & (0.149) & (0.145) & (0.132) \\ 
Harmful (treatment) & 0.303 & 0.400$^{*}$ & 0.753$^{***}$ \\ 
  & (0.191) & (0.185) & (0.169) \\ 
Explainability (treatment) & $-$0.205 & $-$0.053 & $-$0.010 \\ 
  & (0.176) & (0.171) & (0.156) \\ 
Discrimination (treatment) & 0.045 & 0.881$^{***}$ & 0.560$^{***}$ \\ 
  & (0.169) & (0.164) & (0.150) \\ 
Study 2 & $-$0.068 & 0.029 & 0.164 \\ 
  & (0.163) & (0.159) & (0.145) \\ 
Study 3 & $-$0.258 & $-$0.113 & 0.156 \\ 
  & (0.159) & (0.155) & (0.141) \\ 
Harmful:Explainability & 0.205 & 0.039 & 0.023 \\ 
  & (0.261) & (0.254) & (0.232) \\ 
Harmful:Discrimination & 0.268 & $-$0.121 & $-$0.281 \\ 
  & (0.264) & (0.257) & (0.235) \\ 
Intercept & 3.135$^{***}$ & 3.773$^{***}$ & 4.088$^{***}$ \\ 
  & (0.219) & (0.213) & (0.194) \\ 
\midrule
Adjusted $R^2$ & 0.438 & 0.105 & 0.077 \\
Observations & \multicolumn{1}{c}{993} & \multicolumn{1}{c}{993} & \multicolumn{1}{c}{993} \\ [1.8ex] 
\end{tabular} 
  \caption{Regression analysis of blame judgments of the AI system, its developer, and user as a function of participants' explanation of their blame judgments towards the AI system. We report the number of responses coded as each category in the first column. Standard errors are shown inside parentheses. All responses in Study 1 were coded as not harmful, as their mean perceived harmfulness was closer to the perceived harmfulness of Study 2's and 3's not harmful conditions. $^{\dagger}$\pvalue{.1}, $^{*}$\pvalue{.05}, $^{**}$\pvalue{.01}, $^{***}$\pvalue{.001}.} 
  \label{tab:full_regression} 
\end{table*} 